\documentclass{aa}
\usepackage{graphicx}
\usepackage{natbib}
\bibpunct{(}{)}{;}{a}{}{,}
\begin{document}

\title{Are C-rich ultra iron poor stars also He-rich?}
\titlerunning{Are CRUMPS He-rich?}

\author{Georges Meynet$^1$, Raphael Hirschi$^{2,3}$, Sylvia Ekstrom$^1$,  \\
Andr\'e Maeder$^1$, Cyril Georgy$^1$, Patrick Eggenberger$^1$ and Cristina Chiappini$^1$}
\authorrunning{Meynet et al.}

\institute{$^1$ Geneva Observatory, Geneva University, CH--1290 Sauverny, Switzerland. email: georges.meynet@unige.ch\\
$^2$ Astrophysics Group, EPSAM Institute, University of Keele, Keele, ST5 5BG, UK. email: r.hirschi@epsam.keele.ac.uk\\
$^3$ Institute for the Physics and Mathematics of the Universe, University of Tokyo, 5-1-5 Kashiwanoha, Kashiwa, 277-8583, Japan.}

\date{Received  / Accepted }

\offprints{Georges Meynet} 
   
\abstract{The three most iron poor stars presently known ([Fe/H] equal to -5.96, -5.4 and -4.75) are carbon-rich, they are called C-Rich Ultra-Metal Poor Stars (CRUMPS). 
The origin of their peculiar surface abundances is not understood.}
{We propose a synthetic view of the different models so far proposed to explain the peculiar abundances observed at the surface of the CRUMP stars. We deduce some expected trends based on nucleosynthetic arguments and
look for signatures allowing to discriminate among models.
We discuss the conditions for having CRUMP stars which are He-rich,  i.e. with a mass fraction of helium greater than 0.30 and up to 0.60.}
{We discuss the chemical composition of stars made of interstellar medium mixed with wind material of very metal poor massive stars, with wind plus supernova ejecta and with material extracted from the envelope of early AGB stars. Rotating and non-rotating models are considered.}
{The high nitrogen abundances observed in CRUMP stars imply that the material which is responsible for
their peculiar abundance pattern must be heavily enriched in primary nitrogen. We show that rotating stars (both massive
and intermediate mass stars) can produce the required amount of primary nitrogen, and can also account for the
observed enhancements in C, O, Na, Mg and Al. CRUMP stars formed from wind material of massive stars mixed with small amounts of pristine interstellar medium are He-rich
(helium mass fraction between 0.30 and 0.60), Li-depleted and  present low  $^{12}$C/$^{13}$C ratios (inferior to 10 in number).
Such He-rich stars, if discovered, would confirm that the most metal poor CRUMPs formed from essentially pure wind/envelope material.
They would provide the most direct way to probe the nucleosynthetic outputs of the first generations of stars.}
{We show that rotation is a key ingredient to explain the abundance patterns of CRUMPS stars 
 and probably also of at least some Carbon-Enhanced Metal Poor (CEMP) stars, in particuliar the CEMP-no stars. 
Similar non-rotating models, without any extra-mixing, do not succeed to explain the enhancements in the three CNO elements.}  
 
\titlerunning{Are CRUMPS He-rich?}   

\maketitle

\section{Introduction}

The most iron-poor objects found to date in the Universe are halo field stars with [Fe/H] as low as
-5.96 \citep{Frebel2008}. Surprisingly, the three most iron poor  stars present strong overabundances of carbon with respect to iron. 
More generally, \citet{Lucatello2006} find a lower limit of 21$\pm$2\% for the number of stars with [Fe/H]$\le$-2.0 and [C/Fe]$\ge$+1.0.
These stars are collectively named C-Enhanced Metal Poor (CEMP) Stars. The C-rich Ultra Metal Poor Stars (CRUMPS) are a subset of CEMP stars, those with [Fe/H]$\le -4$. 
A more correct name should be C-Rich Ultra  {\it Iron} Poor Stars. The overall metallicity, where the metals would not
only contain iron but all the elements heavier than helium, would amount to a [M/H] above -1, which is, by far,
not ultra metal poor!

For some of these stars,
strong overabundances of nitrogen, oxygen, sodium, magnesium and aluminium with respect to iron 
have been found. Compared to solar values, the abundance ratios may be one to four orders of magnitudes greater
depending on the element and/or star considered.

The CEMP stars can be classified into different categories  according to the presence or not of
{\it s-} and/or {\it r-}process elements  \citep[see the discussion in][]{Masseron2009}.  In the present work, when comparisons are made with observations, we shall focus on the
``CEMP-no'' category, {\it i.e.} on those stars which present no evidence for enhancements in the 
{\it s-} and {\it r-}process elements. 
The absence of {\it s-}process elements  makes less favorable
the possibility that the peculiar surface abundances of these stars result from accretion of material
from an AGB companion, which is in general rich in {\it s-}process elements.
As can be seen in  Fig.  21 of \citet{Masseron2009}, the ``CEMP-no'' category contains the most iron poor objects and are thus
the best candidates to study the enrichment processes due to short lived massive stars. Despite concentrating here on  ``CEMP-no''  stars, the model studied in the present work may also be relevant
for at least some CEMP stars of other categories. 


The main purpose of the present work is to discuss in details
the consequences of the  ``spinstar'' model proposed by \citet{Meynet2006} and \citet{Hirschi2007}
where CRUMPS are formed from material ejected by metal poor rotating stars (hereafter called source material) mixed with some amount of interstellar material. 
We present the various outputs which can be obtained from such models, distinguishing
the following three possibilities for the origin of the source material:
\begin{enumerate}
\item The  wind material ejected by one massive spinstar (spinstar wind model). 
\item Material ejected under the form of wind material during the whole massive spinstar lifetime together
with some amount of matter ejected at the time of the supernova event (spinstar wind and supernova model).
\item The envelope of intermediate mass spinstar at the early AGB phase (spinstar E-AGB model).
\end{enumerate}

We shall discuss the main characteristics which may distinguish these various origins for
the source material.  In particular, we present the conditions required  for producing CRUMP/CEMP-no stars,
which in addition to be C-rich, are also He-rich.

The fact that some CEMP or CRUMP stars may be He-rich have
important consequences for the determination of their physical characteristics:
\begin{itemize}
 \item The mass determinations of the C-rich stars will be different depending on the abundance of helium
considered. Typically, a star, at a given position in the HR diagram, has a smaller mass if it
is He-rich. 
\item Adopting a He-rich model atmosphere modifies the temperature and density structure of the outer layers
where the absorption lines are formed. This may have an impact on the abundances
deduced from spectral synthesis.
\end{itemize}

The discovery of very iron-poor He-rich stars in the field of the halo would be interpreted in the frame of the present model
as the existence of stars made up of nearly ``pure'' source material. Such objects
would therefore represent the most direct opportunities for studying the nucleosynthetic outputs
of the first stellar generations.

He-rich stars are already indirectly found in some globular clusters \citep[see e.g.][]{Piotto2007}.
They may also be present in elliptical galaxies. Let us recall that
all ellipticals show the UV upturn phenomenon, i.e. an UV excess, observed in their spectra.
These excesses are caused by a population of stars on the
blue end of the horizontal branch, {\it i.e.} by 
an old population of hot helium-burning stars \citep[see the review by][]{OConnell1999}.
Different models have been proposed
to explain the origin of this population  \citep[see e.g.][]{Bressan1994, Han2007}.
A possibility, which still remains to be explored, would be that these stars are the descendants
of He-rich stars \citep{Meynet2008c}. He-rich small mass stars are indeed expected to populate the
blue end of the horizontal branch \citep{DecressinIAU2582009}.
 
The paper is organized as follows: 
in Sect. 2, we recall the main characteristics of the ``spinstar'' model. Sections 3  to 5
discuss the chemical
abundance of CRUMPS expected in the frame of the three possibilities listed above. 
In each case, we compare the theoretical expectations
with the observed abundances in the most iron poor stars known today and in CEMP-no stars. 
In Sect. 6 we briefly discuss consequences of alternative scenarios found in the literature. 
The isotopic ratio $^{12}$C/$^{13}$C can be used to discriminate between different models. This
is presented in Sect. 7.
 Conclusions and perspectives are given in Sect. 8.

\section{The ``spinstar'' model}

 \begin{figure*}
  \centering
  \includegraphics[height=12cm, angle=-90, width=17cm]{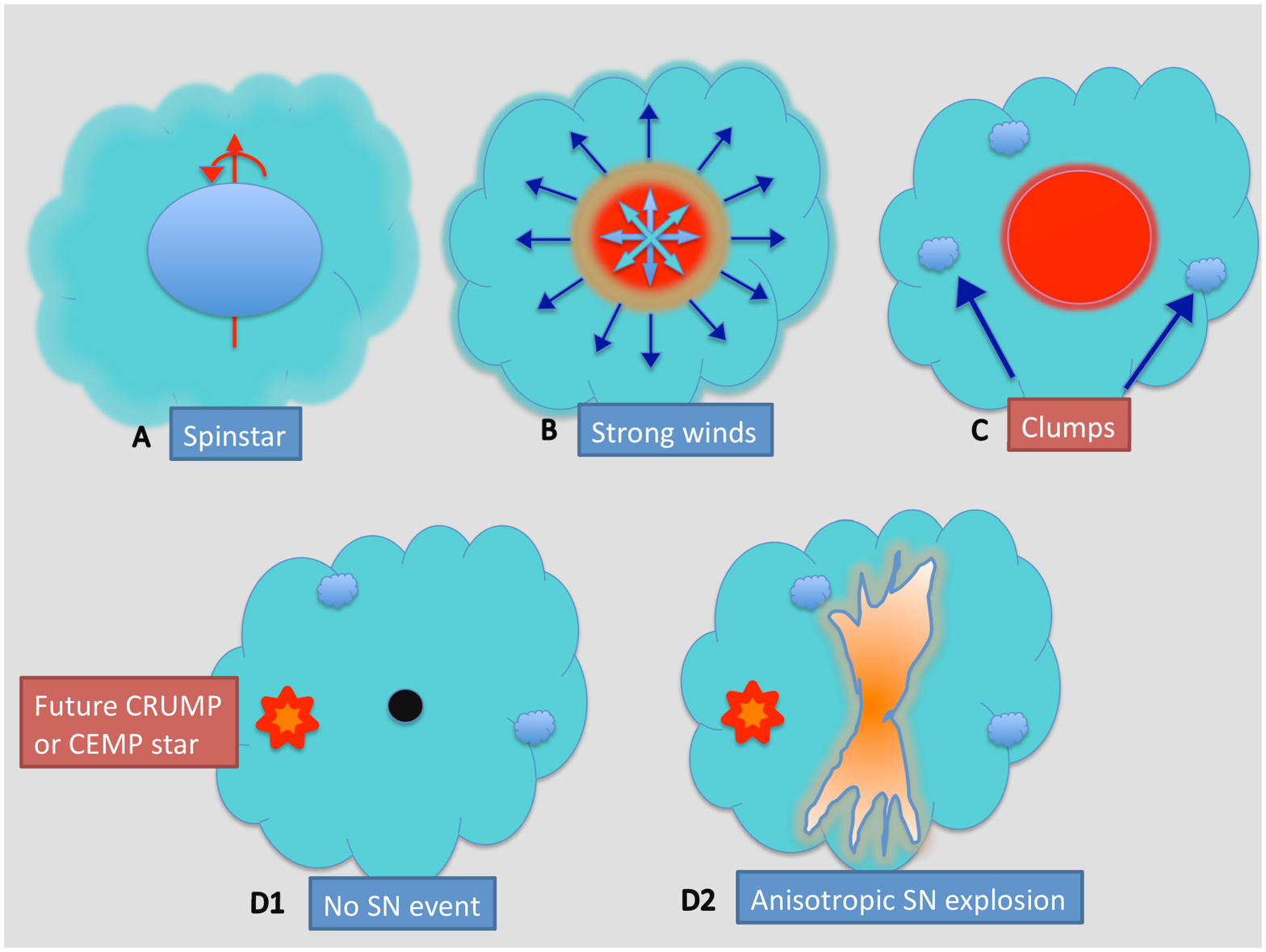}
     \caption{Schematic representation of the ``spinstar'' model for explaining the origin of the peculiar abundance patterns of CRUMP/CEMP-no stars: A) A metal poor massive rotating star  evolves during the core H-burning phase.  B) During the core He-burning phase,  strong winds,  triggered by self enhancement of the surface metallicity appear .  C)  Wind material mixed with some amount of  interstellar medium form clumps. The end of the evolution of the spinstars can either be a Black Hole with no supernova explosion (D1) or it can explode as a supernova (D2) which can further enrich the clumps. The fast rotation of the core may  produce a non isotropic explosion. Some clumps form small mass stars, which are the CRUMP/CEMP stars that we can observe today.
             }
        \label{schema}
  \end{figure*}
  
The term ``spinstar'' refers to those
very metal poor stars which rotate sufficiently fast for having their evolution strongly affected by rotation
\citep{Meynet2008a}. 
Spinstars can be either massive or intermediate mass stars.

{\it Rotation} is needed for two purposes, first to allow mixing between the He- and the H-burning regions
and thus the synthesis of important amounts of primary $^{13}$C,  $^{14}$N and $^{22}$Ne. Rotation also
deeply affects the {\it s}-process element nucleosynthesis \citep{Pignatari2008}. Second, rotation is important
to trigger important mass loss episodes occurring during the evolution of these very metal poor stars.
This is due again
to rotational mixing, which transports primary elements to the surface. This increases  the opacity
of the outer layers and thus stimulates mass loss by line driven winds.

{\it Very metal poor environment ($Z$ below 0.001)} is needed because rotational mixing is more
efficient at low metallicity  \citep{Maeder2001}.  In addition,
on average, stars seem  to rotate faster at low metallicities \citep[see e.g. the review by] [and references therein] {Meynet2008b}.


The effects discussed above are consequences of the same physics which, when applied to more familiar metallicities, is able to reproduce successfully many observed features such as
the changes of the surface abundances \citep{Maeder2009a},
the correct number ratio of blue to red supergiants at low metallicity \citep{Maeder2001},
the variation with
the metallicity of the number ratio of Wolf-Rayet  (WR) to O-type stars, 
the existence of WR with surface
abundances rich in both H- and He-burning products,
 the correct WN/WC number ratio at low metallicity \citep{paperXI2005}, 
 the variation with the metallicity of the ratio of type Ib and Ic to type II supernovae \citep{Georgy2009}. 

With respect to the models reproducing the above observed trends,
just two ingredients have been taken
differently, obviously the metallicity which was chosen very low ($Z$ = 10$^{-8}$ and 10$^{-5}$) and the initial ZAMS
rotational velocity on the equator. 
At such low metallicities, there is no guideline for choosing the initial rotational velocity,  except that it should be inferior
to the critical velocity ({\it i.e.} the velocity at which the centrifugal acceleration at the equator
compensates for the local gravity).
We chose ratios of $\upsilon_{\rm ini}/\upsilon_{\rm crit}$ between 50 and 70\% on the ZAMS depending
on the initial mass and metallicity considered (see Table \ref{data}). These values are well below 1 but above the usual ratio adopted for
solar metallicity models, which is 40\%.


The general qualitative outline of the spinstar scenario  is shown in Fig. \ref{schema}. In 
A) a metal poor rotating star  evolves during the core H-burning phase. It may happen that
     the surface velocity reaches the critical velocity during this phase forming an equatorial disk enriched in H-burning products. At the very low metallicity considered here however, the amount of mass lost in that way is quite modest. B) During the core He-burning phase, as a result of rotational and convective mixing which brings to the surface primary CNO elements, the opacity of the outer layers is increased. The star evolves to the red part of the HR diagram. A deep
     outer convective zone appears, dredging up large quantities of CNO elements at the surface. This triggers strong mass loss. The wind is made of both H- and He-burning products; C) Part of the wind material forms clumps (some amount of interstellar medium may enter into their composition, see below). If stars form later in these clumps, their initial composition will bear the mark of the composition of the spinstar's ejected material (source material).
  Sufficienly long lived small mass stars formed in that way produce the CRUMP stars that we observe today.
  
 The end of the evolution of the spinstars, if massive, can be a black hole with no supernova explosion (D1). In that case there is no risk
 that the clumps be destroyed or further enriched by the supernova ejecta. The massive spinstar might also explode as a supernova (D2). In this case, the fast rotation
 of the core may  produce a non isotropic explosion \citep{Tominaga2007a} which may allow some clumps to survive the explosion. 
 More probably, an intermediate situation will arise where a faint supernova explosion occurs with a lot
 of material falling  back on the nascent black hole as has been proposed by \citet{Umeda2003}.
 
Enrichment of some clumps by supernova ejecta may occur.
Therefore, in this general frame, various enrichment scenarios are possible  depending on the nature of the spinstar,
on the degree of mixing with interstellar material, and on the importance of the wind contribution 
with respect to the supernova one.

 \begin{figure}
  \centering
  \includegraphics[height=11cm, width=9cm]{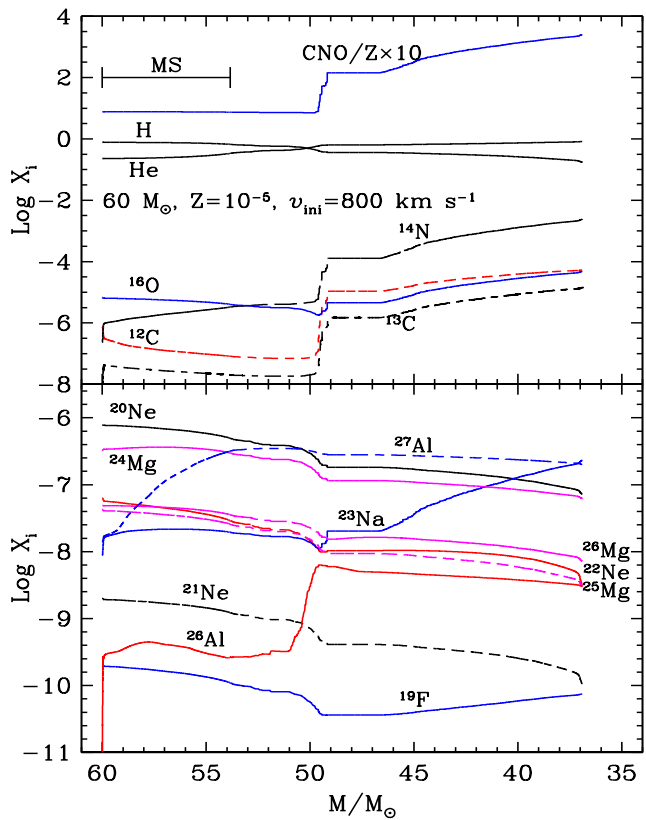}
     \caption{Evolution of the surface abundances for a 60 M$_\odot$ model with $Z=10^{-5}$ and 
     $\upsilon_{\rm ini}=800$ km s$^{-1}$
     (model E in Table \ref{data}) as a function of the actual mass of the star. 
     The actual mass decreases as a function of time. The extension of the Main Sequence (MS) phase is
     indicated in the upper panel. The curve labeled CNO/Z$_{\rm ini}$  shows the evolution of the (logarithm of the)
     ratio ($^{12}$C+$^{14}$N+$^{16}$O)/Z at the surface. The curve has been shifted to 1 dex upwards for clarity.
     During the MS phase, the sum of the CNO element remains constant. At a given point, corresponding
     to an actual mass of about 50 M$_\odot$, the sum of CNO elements increases strongly as a result of rotational
     mixing in previous phases and the deepening of the outer convective zone. At that stage the star has an effective
     temperature of the order of 3.85.
             }
        \label{fig1wind}
  \end{figure}

\section{The spinstar wind model}

In Fig. \ref{fig1wind}, we show how the surface abundances evolve as a function of the actual mass of the star (which
decreases as a function of time) for a 60 M$_\odot$ model at $Z$=0.00001 and with $\upsilon_{\rm ini}/\upsilon_{\rm crit}=0.72$.
During the MS phase, the star
loses only a little more than 6 M$_\odot$. The material ejected during the MS phase bears the usual signature
of the CNO cycle, {\it i.e.} increase of $^{14}$N and decreases of $^{12}$C, $^{16}$O,  and of the $^{12}$C/$^{13}$C 
ratio. We note also the consequences at the surface of the action
of the Ne-Na and Mg-Al chains which are active in the H-burning core (decrease of Ne and Mg isotopes and increase
of $^{23}$Na and $^{27}$Al).
At a given point (on Fig. \ref{fig1wind} when the actual mass is around 50 M$_\odot$), He-burning products, as  $^{12}$C and
$^{16}$O  appear at the surface,
while those due to H burning, as e.g. nitrogen, are also increased. Rotational diffusion  transports He-burning products in the H-burning shell
where part of them are transformed by the CNO processing  (this explains
the strong $^{13}$C and $^{14}$N enhancements). Part of the He-burning products succeed in diffusing through the H-burning shell up to the surface without being
destroyed by the CNO cycle (this explains the strong $^{12}$C and $^{16}$O abundances).



\begin{table*}
\caption{Masses ejected by stellar winds from different rotating models and $^{12}$C/$^{13}$C  number ratios in the ejecta.
}
	\centering
		\begin{tabular}{clccccccccccc}
		\hline
		\hline 
Model &M$_{\rm ini}$  & Z$_{\rm ini}$  & $\upsilon_{\rm ini}$       & $\upsilon_{\rm ini}/\upsilon_{\rm crit}$   &     M$_{\rm eje}$ & M$_{\rm H}$ & M$_{\rm He}$ 
& M$_{\rm C12}$ & M$_{\rm C13}$       & M$_{\rm N14}$ & M$_{\rm O16}$  &  $^{12}$C/$^{13}$C      \\
            & [M$_\odot$] &                               & [km s$^{-1}$]                    &                                                                    & [M$_\odot$]     &  [M$_\odot$]     &  [M$_\odot$]  &
  [M$_\odot$]  &  [M$_\odot$] & [M$_\odot$]  &  [M$_\odot$] &  \\           
	  \hline 
A & 85                 & 10$^{-8}$ & 800 &  0.53 &             65.2  & 17.9 & 35.6         & 6.3             & 0.6             & 1.7            & 3.0             &  11.4 \\
B &60                 & 10$^{-8}$ & 800 &   0.57 &           11.0  &   7.2 &   3.8          & 0.000020 & 0.000005 & 0.000690 & 0.000050  &    4.3 \\
C& 60                 & 10$^{-8}$ & 800 &   0.58  &  36.2  & 14.5 &  21.5        & 0.005        & 0.001        & 0.2             & 0.006        &    5.4 \\
D& 40                 & 10$^{-8}$ & 700 &    0.55 &            4.2  &   2.9 &   1.3          & 0.005        & 0.0008      & 0.004        & 0.002        &    6.8 \\
	&	   &       &    &  & & & & & &   & & \\
E& 60                 & 10$^{-5}$ & 800 & 0.72    &         23.1  & 10.9 & 12.2         & 0.0003      & 0.00007   & 0.010         & 0.003        &  4.6 \\
F&  7                  & 10$^{-5}$ & 450 &    0.64     &       5.7  &    3.6 &   2.0         & 0.05           & 0.0007     & 0.02           & 0.004        &  77.4\\
	\hline \\	   
		\end{tabular}
		\label{data}
\end{table*}
   
In Table \ref{data}, we indicate for the various rotating models computed at very low
metallicity the following quantities:  columns 2 to 5 give respectively
the initial mass, metallicity, the initial rotation velocity on the ZAMS, $\upsilon_{\rm ini}$, and the ratio $\upsilon_{\rm ini}/\upsilon_{\rm crit}$.
The total mass ejected by the winds is given in column 6. The total masses ejected under
the form of hydrogen, helium, $^{12}$C,  $^{13}$C,  $^{14}$N and  $^{16}$O are given in columns
7 to 12, respectively. The number ratio $^{12}$C/$^{13}$C is indicated in column 13. The models for $Z$ =10$^{-8}$  ([Fe/H] = -6.6) 
are from \cite{Hirschi2007} except for one, model C,   taken from \cite{Meynet2006}.
Models at 10$^{-5}$ ([Fe/H] = -3.6) are from \cite{Meynet2006}.
Note that some differences in the input physics explains the different outputs for otherwise similar models  (see the above references for more details). 

From Table \ref{data}, it is easy to compute the initial chemical abundance that a star would have if formed
 from a mixture of wind material  and of some amount of pristine interstellar matter. If $M_{\rm wind}$
 is the mass ejected under the form of wind and $M_{\rm ISM}$ the mass of interstellar material with which
 the wind material is mixed, then let us define $D$, the dilution factor, as the ratio $M_{\rm ISM}/M_{\rm wind}$.
 Then the mass fraction of element $i$, $X_{i}$, in the star formed from wind ejecta and interstellar matter is given by
 $$X_{i}={{M_i \over M_{\rm wind}}+X_{i0} \times D \over 1+D},$$
 where $M_i$ is the mass of element $i$ ejected by the winds,
 $X_{i0}$ is the mass fraction of element $i$ present  in the
 interstellar material.
The quantity $[X_{i}/{\rm H}]$ can be computed from the formula below
$$
[X_{i}/{\rm H}]=\log
\left(
{
{M_i \over M_{\rm wind}}
+X_{i0} \times D 
\over
{M_{\rm H} \over M_{\rm wind}}+X_{\rm H0}\times D }
\right)
-\log
\left(
{X_{i\odot} \over X_{\rm H\odot}}
\right)
,$$ 
where $X_{i\odot}$ is the mass fraction of element $i$ in the Sun\footnote{For the quantities 
of the type [A/B] the result is the same using mass fraction or number fraction provided of course
the same type of quantities is used to compute the elemental abundances in the star and in the Sun.}.
We used as interstellar abundances the initial abundances
used to compute the models \citep[][]{Meynet2006, Hirschi2007}. The solar abundances are those of \citet{Asplund2005}.
Note that these solar abundances are the same as those used for plotting the observed [Xi/H] ratios in HE 1327-2326
(see e.g. the values plotted as star symbols in Fig. \ref{obsdata}).

To compute the expected chemical composition of CRUMP stars made of such wind material, a first step
consists in obtaining information on the dilution factor. This is the topic of the next subsection. 


\subsection{Constraint on the dilution factor}



An interesting element to be used for constraining the dilution factor is lithium. 
It was used by \citet{Decressin2007a}
for estimating the dilution factor in another context, namely that of the chemical anomalies observed
in globular clusters. We follow here the same line of reasoning
in the context of the CRUMP stars found in the field of the halo.

Lithium has a reasonably well known value in the pristine interstellar medium equal to
the value given by standard Big Bang nucleosynthesis. It is completely destroyed in massive stars and also in AGB stars.
Thus any mixing
of such stellar ejecta with pristine interstellar material will increase the abundance of lithium with respect to the abundance in the source material.
The dilution to be adopted is the one which will allow to reproduce the observed lithium abundance.
Of course, this is correct as long as the abundance of lithium is not affected by internal mixing processes
having occurred in the CRUMP star itself.  

For the star HE 1327-2326 ([Fe/H] $\sim -5.96$),  \citet{Frebel2008} obtained an upper value
of $\epsilon$(Li)  of 0.62, with $\epsilon$(Li) defined as
12+$\log(N({\rm Li})/N({\rm H}))$, where $N$(Li) and $N$(H) are the density number
of lithium and hydrogen. This is a very low value, both compared to the abundance
of Li expected from standard Big Bang nucleosynthesis
\citep[2.72 according to][]{Cyburt2008}, and to the value of Li observed in the bulk of normal metal poor halo stars on the Spite plateau 
\citep[2.10 according to][]{Bonifacio2007}.

As already indicated above, HE 1327-2326 is either an evolved main-sequence or an early subgiant star with an effective temperature of about 6200 K.
Although there is no ``normal stars''  ({\it i.e.} not C-enhanced) observed with such a low iron content and thus no possibility to
know the level of the Spite plateau in this range of iron-content, such a star presents at least all the characteristics
(evolutionary stage and effective temperature) for belonging to the Spite plateau. Thus one would have expected
a value around 2.10 instead of an upper limit of 0.62. 

Recently \citet{Korn2009} computed the effects of atomic diffusion
including gravitational settling, thermal diffusion and radiative accelerations in models with masses around 0.78 M$_\odot$ and an
iron content similar or lower than that of HE 1327-2326.
In one model, an additional ad hoc diffusion coefficient is added with no attempt to connect it to a physical process like rotation.
This diffusion coefficient was chosen to reproduce the observed abundance trends between stars from the turn off point to the red giant
branch in NGC 6397 \citep{Korn2007, Lind2008}. For what concerns lithium, the model with no additional diffusion shows a depletion of
1.2 dex and the one with additional diffusion shows a depletion by 0.3 dex. Thus we  see that, according to this study, depletion mechanisms,
would not succeed in decreasing the lithium abundance from the cosmological value below the observed upper limit. Indeed to obtain such a result
a depletion of at least 2.1 dex is required! Moreover the case with no additional mixing, which produces the greatest depletion factor, is probably not the favored one since it
cannot explain the observed trends in NGC 6397. From this we conclude that at the present time there is no strong reason to suppose
that the very low lithium abundance observed today in HE 1327-2326 results mainly from internal depletion processes.
Thus we are left with the second possibility, that the low Li-abundance observed today mainly results
from the fact that this star was formed from very Li-poor material.

 \begin{figure}
  \centering
  \includegraphics[height=12cm, width=9cm]{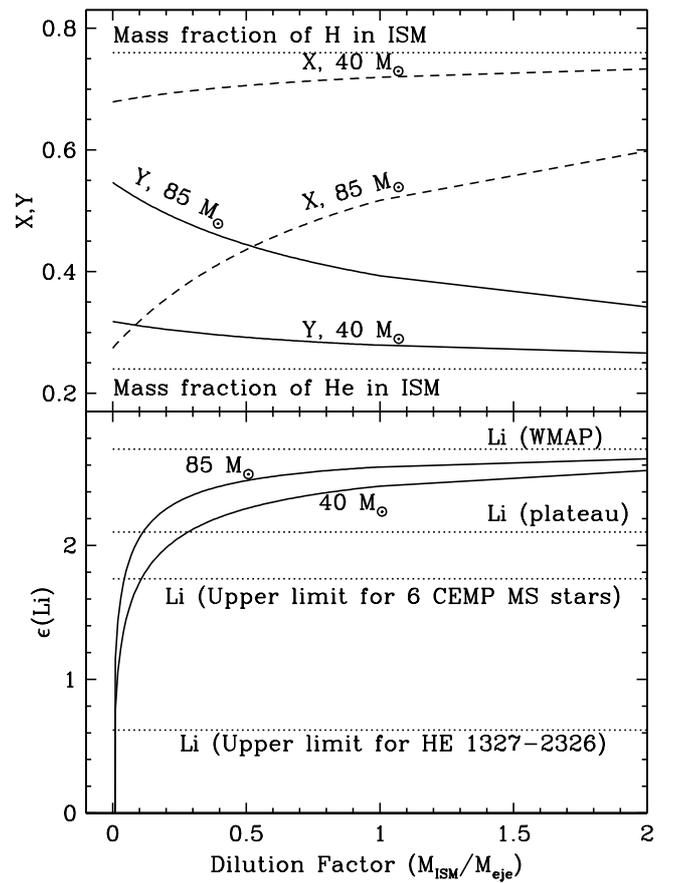}
     \caption{Predicted abundances in stars made of a mixture of wind ejecta of massive rotating stars
     and of pristine interstellar medium plotted as a function of the dilution factor $D$, the ratio between the mass
     of interstellar medium and the mass 
     ejected in the wind involved in the mixture. The rotating models
     are for 40 and 85 M$_\odot$ models at $Z=10^{-8}$. The 85  M$_\odot$ model has a rotational velocity on the ZAMS equal to 800 km s$^{-1}$. The initial velocity for the 40  M$_\odot$ model is 700 km s$^{-1}$.
     In the upper panel, the mass fractions of hydrogen (dashed lines) and helium (continuous lines) are shown. The dotted
     lines indicate the mass fraction of these two elements in the interstellar medium. The lower panel shows the
     variation of the lithium abundance as a function of the dilution factor  for the two initial masses considered
     (continuous lines and see text for the meaning of $\epsilon$(Li)). The WMAP cosmological abundance of Li is indicated 
     by a dotted line, also the value of the Li abundance of the Spite plateau, an upper limit for the Li abundance observed
     in Main Sequence CEMP stars, and the upper limit observed in HE 1327-2326   (see references in the text). 
             }
        \label{figdilu}
  \end{figure}

In the bottom panel of Fig. \ref{figdilu}, the continuous lines show how the abundance of Li varies in a body formed from
spinstar winds (completely  Li-free) and of pristine interstellar material having WMAP Li abundance
when the dilution factor is increased. The two curves corresponding to our 40 and 85 M$_\odot$ models do not
superpose because the wind ejecta of these two models do not have the same mean hydrogen abundance and $\epsilon$(Li)
show the variation of the abundance of lithium with respect to hydrogen (see the definition just above).

Let us consider the case of the 40 M$_\odot$ model. In case no Li-depletion at all occurred in the CRUMP star,
then the present day observed value is equal to the initial value. In that case, Fig. \ref{figdilu} tells us that the star
should be made from pure ejecta. Any dilution would rapidly bring the Li abundance above the observed upper limit.

In case the depletion factor is 1.2 dex  \citep[maximum and probably unrealistic value inferred from the models of][]{Korn2009}, then the observed upper limit would be compatible with
a dilution factor inferior to about 0.13 (look at the abscissa of the point on the 40 M$_\odot$ model with an ordinate equal to 0.62+1.2=1.82).
Although not zero, the dilution factor remains extremely small.

We have also indicated in the bottom panel of Fig. \ref{figdilu}  the upper limit abundance measured in 6 CEMP Main Sequence stars
as obtained from the SAGA database of \citet{Suda2008}. Note that there are only 7 stars in total with measured Li abundance and low [Fe/H].
Except for one star, which is on the Spite plateau, the other 6 stars in this sample all have values of Li inferior to 1.75. 
Note also that the two stars in this sample with [Fe/H] $<$ -4 have an upper limit for the Li abundance of 1.5 and 1.1 respectively.
The upper limit of 1.75 is compatible with a dilution factor of 0.11 and 0.25  if, respectively, there has been no 
Li- depletion or a 0.3 dex depletion in these CEMP stars. In case depletion of 0.6 dex would be allowed  (difference between the WMAP and Spite Plateau) then the dilution factor can be at most 0.65.

From the discussion above we conclude that for reasonable value of Li-depletion having occurred in the CRUMP star itself, 
then the low values of Li observed in some of  the CEMP stars can only be explained if two conditions are fulfilled:
1) Low Li-CEMP stars are made of Li-free (or nearly free) material; 2) a small amount of pristine
interstellar medium can be added, at most 65\%.

Interestingly, this means that the appropriate abundances for the low Li- CEMP stars must be obtained mainly from the source. The dilution factor cannot be 
used as a free parameter allowing to adjust the level of the abundances.
This implies also that all the potential sources which are able to produce Li are discarded as
candidates for providing low Li-CEMP stars material. This is not however a strong constraint since
all the potential candidates considered so far for explaining the origin of the CEMP star material,
namely the Asymptotic Giant Branch (AGB) stars, the supernovae and of course the spinstars provide Li-free material\footnote{Li may only be synthesized at some peculiar phases in stars, as
in the red giant phase of low mass stars through the Cameron-Fowler mechanism \citep{Cameron1971}. It can also be produced
by spallation reactions induced by galactic cosmic ray reaction with the interstellar medium \citep{Meneguzzi1971}.}. 
  
\subsection{Consequence of the small dilution factor: some CEMP stars are He-rich}

In the upper panel of Fig.\ref{figdilu}, we show  how the abundance of helium  (in mass fraction) varies in a body formed from
spinstar winds  and of interstellar material with a mass fraction of helium equal to 0.24
when the dilution factor is increased. 

As can be seen from Table 1, the masses ejected by both the 40 and 85 M$_\odot$ models are  sufficient for forming at least
one 0.8 M$_\odot$ stars. Such a star, made of a  pure rotating 85 (40)  M$_\odot$ star wind ejecta would have  an initial $Y \sim 0.55$ (0.32).
For any dilution  factors inferior to  about 4.1 (0.3), a CEMP star made of wind ejecta of a
85 (40)  M$_\odot$ model,  would have a helium mass fraction superior to 0.30. 

This conclusion can be formulated in a more general way: any stars formed mainly from the envelope
of an evolved massive or intermediate mass star,  rotating or not, would be He-rich \citep[see Table 2 in][]{MeynetIAU265}.
Thus He-richness would not be per se a hint that the spinstar model we propose here is correct
but that the CRUMP stars are mainly formed from the envelope of evolved stars.

Would He-rich stars be necessarily Li-poor? The answer is no.
The example above shows that even with a dilution factor of 4,
(see the case of the 85 M$_\odot$ model), we still have a He-rich star, while the initial Li abundance, obtained with such
a dilution factor, would be nearly equal to the
cosmological value! 
Are Li-poor stars necessarily He-rich? No because the CRUMP star may be formed
from a high fraction of interstellar medium and lithium may have been then depleted
by internal mixing processes in the CRUMP star itself, although this is unlikely to happen in current stellar evolution models.



\subsection{Predictions for CNO abundances in CRUMP/CEMP-no stars}

\begin{figure}
  \centering
  \includegraphics[height=9cm, width=9cm]{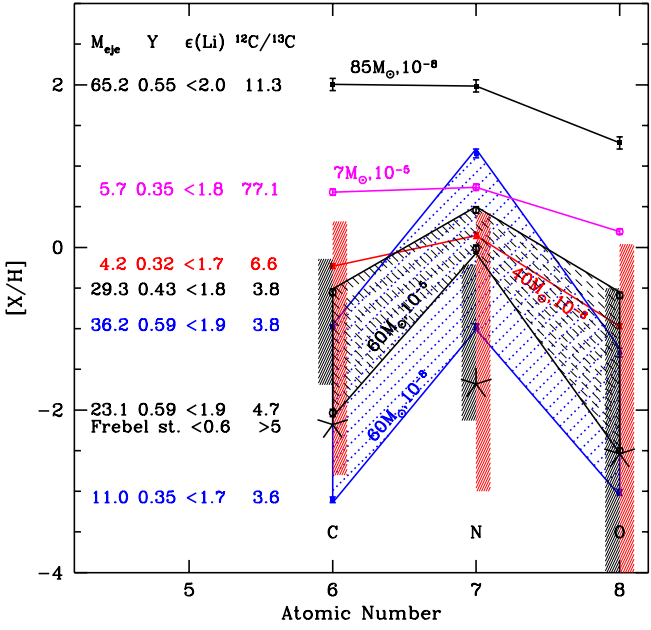}
     \caption{The [C/H], [N/H] and [O/H] ratios in the wind ejecta of various rotating models mixed with at most 15\% of pristine interstellar medium are shown (dilution factor $D=0.15$). The models are those
     presented in Table \ref{data}.  The models obtained for the same initial mass, metallicity and rotation but computed with
     different prescriptions cover the shaded areas.
     The vertical hatched columns show the range of observed values for CEMP-no stars with log $g$ superior to 3.8 (left)
     and with log $g$ inferior to 3.8 (right).  The values were taken from Tables 1 and 2 of \citet{Masseron2009}. Only upper limits are given for [O/H], this is why the columns extend down to bottom.
     The star symbols show the observed values for the most iron poor star known today (Frebel et al. 2008). 
     On the left part of the figure, are indicated for each models the total mass ejected in solar masses, the mass fraction of helium
    (case with D=0), the upper mass limit for Li abundance (obtained with D=0.15), and the isotopic ratio $^{12}$C/$^{13}$C in number.
             }
        \label{obsdata}
  \end{figure}

In Fig. \ref{obsdata}, we show the values of the [C/H], [N/H] and [O/H] ratios  obtained in the wind ejecta of our rotating models
for dilution factors between 0 and 0.15. 
Except for the 85 M$_\odot$ (see error bars for the values of the 85 M$_\odot$ model), the values obtained with D=0 and 0.15 are identical. 
The helium mass fractions are indicated for a dilution factor equal to 0. 
The upper value of the abundance of Li is given using D=0.15. The ratio  $^{12}$C/$^{13}$C is insensitive to such small variations
of the dilution factor (see Sect. 7).

In Fig. \ref{obsdata}, the vertical hatched zones show the range of observed values for CEMP-no stars with log $g$ superior to 3.8 (left dark grey column, 5 stars in the sample)
     and with log $g$ inferior to 3.8 (right light grey column, 20 stars in the sample, red in the online version).  The observed data were taken from Table 2 of \citet{Masseron2009}.
     The star symbols show the observed values for the most iron poor star known today \citep[HE 1327-2326,][]{Frebel2008}. 
     The second \citep[HE 0107-5240,][]{Christlieb2004} and third \citep[HE 0557-4840,][]{Norris2007} most metal poor star known today belong to the group 
     of the CEMP-no stars with log $g$ inferior to 3.8. These last two stars are giants, therefore their surface abundances may have been
     affected by the first dredge-up.

 \begin{figure}
  \centering
  \includegraphics[height=9cm, width=9cm]{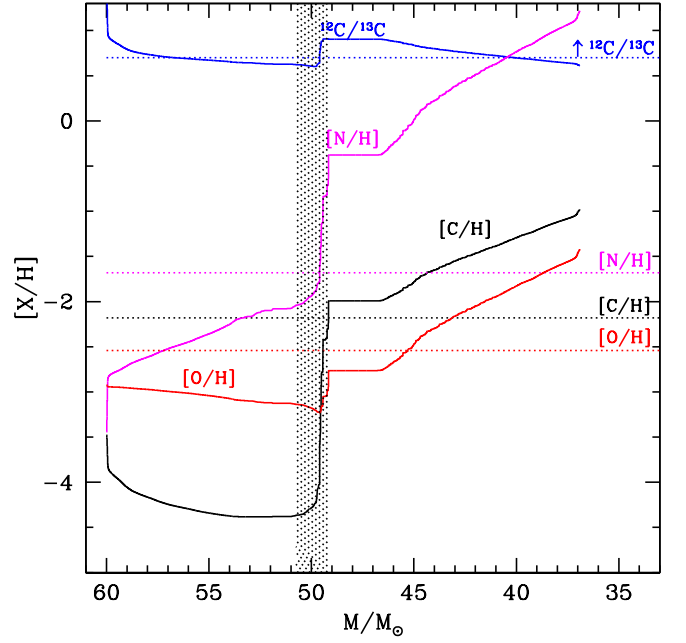}
     \caption{Same as Fig.\ref{fig1wind} for the [X/H] ratios. The curve labelled $^{12}$C/$^{13}$C shows the
     evolution of the number ratio of the two carbon isotopes. The horizontal dotted lines show the values
     obtained for HE 1327-2326 \citep{Frebel2008}. In the case of the carbon isotopic ratio, only a lower
     limit is given. The shaded area represents the portion of the mass ejected under the form of stellar winds
     which shows, once well mixed,  the chemical composition plotted as the third curve from bottom on
     left panel of Fig. \ref{blobZ5}. 
             }
        \label{surch}
  \end{figure}

Before discussing the comparisons with the observed abundances in HE 1327-2326 and CEMP-no stars let us note a few general
trends of the theoretical ratios:
\begin{itemize}
\item  In the solar mixture, the CNO elements represent a number of atoms equivalent to about 80 times the number of iron atoms,
and the number of nitrogen atoms corresponds to about 3 times the number of iron atoms. Let us suppose that these
proportions would be approximately the same at all the metallicities (this is the hypothesis we did for determining the initial
distribution of the heavy elements in our model stars).  In that case what would be the maximum [N/H] value
one might expect if all the carbon and oxygen initially present in the star were transformed into nitrogen? It would be equal to
[N/H]$_{\rm max}$=$ \log$(80/3)+[Fe/H]=1.4 +[Fe/H]. Any [N/H] value shifted upwards by more than about 1.4 dex with respect
to [Fe/H] is an indication that nitrogen has to be produced in a primary process. This happens in all the computed models
shown in Fig. \ref{obsdata}.
\item Primary nitrogen can only be produced in a star which has simultaneously
a H- and He-burning region.  In addition,
some mixing process must occur between these two nuclear active regions.
Rotational mixing is the mechanism we propose here to trigger
this mixing.
As recalled above, the mechanism which allows the ejection
of that material is the self enrichment of the surface, also triggered by rotation. Thus  {\it rotational mixing, which produces the peculiar abundance
pattern, also allows the chemically peculiar matter to be ejected by the star.}
\end{itemize}

\begin{figure*}
  \centering
  \includegraphics[angle=-90, width=15cm]{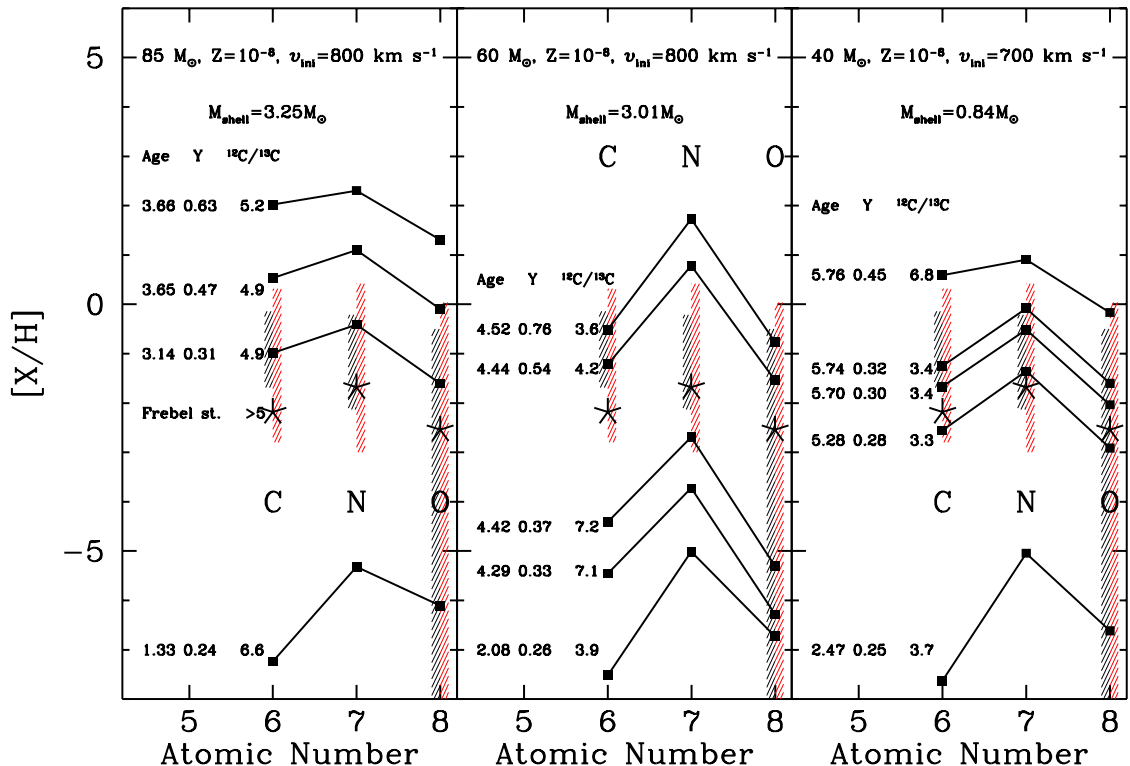}
     \caption{The [C/H], [N/H] and [O/H] ratios in  wind shells ejecta of various rotating models are shown (see text). The models plotted
     are the models A, C and D of Table \ref{data}. 
     The masses of the shells are indicated as well as the age of the star when half of the mass of the  shell is ejected, the mass fraction of helium and the 
     $^{12}$C/$^{13}$C in number  in the shells are indicated  (case with $D$=0).
     The stars show the observed values for the most iron poor star known today \citep{Frebel2008}. The vertical hatched zones have the same meaning
     as in Fig. \ref{obsdata}.
             }
        \label{blobc3}
  \end{figure*}
  
Let us now see how these models compare with the observed abundances in HE 1327-2326. First let us 
note that this star presents strong signs of being made of material having been enriched by a very important quantity
of primary nitrogen. Indeed the [N/H] value of the order of -1.5 is well above the value of -4.56=-5.96+1.4! There are more than 3 dex
of difference. All the present models show very strong productions of primary nitrogen with [N/H] values shifted upwards by
2 to 4 dex with respect to [Fe/H].

We see that the present models provide trends in qualitative
agreements with the observations of  HE 1327-2326: low Li abundances, high [X/H] ratios for simultaneously C, N and O,
low  $^{12}$C/$^{13}$C (although only an upper limit is given for HE 1327-2326). All the models also
predict  He-rich (Y $\ge 0.30$) stars.
None of the present models however fits perfectly all the observed data.
But here three facts have to be recalled:
1) the models are not at all adjusted in order to fit the observed abundances. We have simply plotted
our predictions for the wind ejecta composition. As explained above the dilution factor does not offer any
degree of freedom; 2) changes of the initial rotational velocities, of the initial masses, and probably,
although to a lesser extent, of the initial metallicity would enlarge the range of values predicted by the models; 
3) depending on the evolutionary stage at which
the mass is ejected by stellar winds, its chemical composition can vary a lot. 
This is the point we discuss in the next subsection.
 
Comparing the predictions of our spinstar wind models with the observations of CEMP-no stars
(see hatched columns in Fig. \ref{obsdata}), we see that
globally the situation is rather encouraging. We see that many models cover
the hatched regions.

\subsection{The spinstar shell model} 

In this section we explore a slightly different version of the spinstar wind model exposed in the previous section. In this model the CRUMP stars are still formed from
spinstar ejecta but now, instead of considering the well mixed material ejected by winds during the whole stellar lifetime,
we consider shells of material ejected at different evolutionary stages. Note that we still suppose that stellar winds are continuous but
we consider that only wind material emitted during an interval of time is used to form the CRUMP star. 

Let us begin by examining at which stage the wind ejected material presents the greatest similarities with the abundances
observed at the surface of HE 1327-2326.
In Fig. \ref{surch} the evolution of the [X/H] ratios  and of the number ratio $^{12}$C/$^{13}$C are shown at the surface of the 60 M$_\odot$ model
with $Z=10^{-5}$ (model E in Table 1).  
The $^{12}$C/$^{13}$C is very low most of the time (below 10), while the initial ratio used for computing the models is above 75. However its evolution is
non monotonic. It first decreases, as a result of the CNO cycle, then when He-burning products appear, a rapid increase of the ratio occurs
(around 50 M$_\odot$ in Fig. \ref{surch}).  Then due to the continuous action
of the H-burning shell it decreases again. 

In Fig. \ref{surch} we have overplotted the level of the observed ratios in HE 1327-2326 as given by \citet{Frebel2008}.
The chemical composition of the wind has the greatest similarity with the abundances observed in the Frebel star when
He-burning products begin to appear at the surface. 
Before that time, the matter would be depleted in carbon and oxygen.
After that time, the enhancements of the CNO elements would be too high. 
Of course the evolution shown in Fig. \ref{surch} 
changes when the initial mass of the star changes or when the initial velocity is varied. 

In contrast with Fig. \ref{surch}, where instantaneous values of the surface
composition are plotted, in Fig.  \ref{blobc3}, the composition in shells of material is shown. For instance,
for our 85 M$_\odot$ star, 
we consider 20 events of mass loss, of 3.25 M$_\odot$ each (summing a total of 65 M$_\odot$  lost during the life of this star).
Only the compositions of four or five shells are shown in each panel of Fig. \ref{blobc3}.
The lower curve corresponds to the first shell ejection and the upper one to the last one.
When evolution proceeds, the [X/H]'s shift upwards.

From Fig. \ref{blobc3} the following remarks can be made:
\begin{itemize}
\item We note that already when the first shells are ejected the $^{12}$C/$^{13}$C ratio is quite low. 
It  takes values near the one expected from CNO equilibrium.
This comes from the fact
that this ratio is one of the first to change at the stellar surface. This is true for all the models discussed here. 
\item  When evolution proceeds, some $^{12}$C produced in the He-burning region does appear at the surface and
increases the $^{12}$C/$^{13}$C ratio to values above the equilibrium value. In further evolutionary stages, the continuous action of the
H-burning shell produces again a decrease of this ratio. Thus, in the framework of the spinstar model, we see that {\it a value of
the $^{12}$C/$^{13}$C ratio slightly above the equilibrium CNO value, as observed in Frebel star, 
favors material ejected at the transition between the phase
when only H-burning products are ejected and that when both H- and He-burning products are ejected.} This confirm what
was deduced considering Fig. \ref{surch}.
\item As time goes on, matter processed by H- and He-burning are ejected and the CNO abundance ratios are enhanced
by many order of magnitudes, as well as the mass fraction of helium. 
\item In all three cases there are shells which would fit 
the observed CNO abundances for the Frebel star (see e.g. the shell
ejected at a mean age of 5.28 My by the 40 M$_\odot$ stellar model). Similar cases might be found for the other models, changing
the mass of the shells and/or the time interval during which it is ejected. 
\end{itemize}

\begin{figure}
  \centering
  \includegraphics[width=6.5cm, angle=-90]{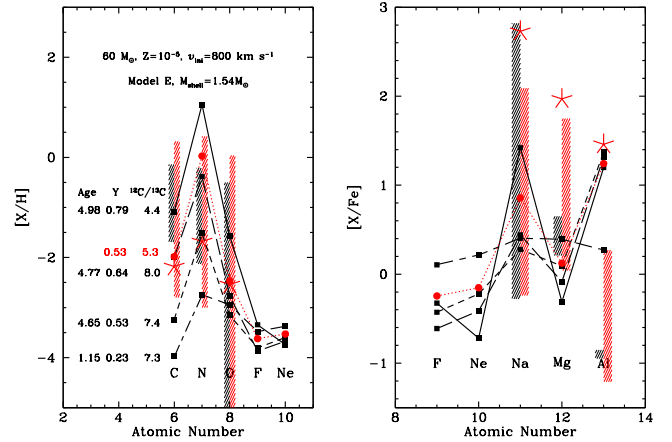}
     \caption{{\it Left panels}: The [C/H], [N/H], [O/H], [F/H] and [Ne/H] ratios in different shells wind-ejected by models E
 of Table \ref{data}. On the left part of the left panel, are indicated for each shell the mean age at the time of shell ejection, the mean mass fraction of helium and the
isotopic ratio $^{12}$C/$^{13}$C in number (case with D=0) in the shell.  {\it Right panels}: The [X/Fe] ratios are shown for $^{19}$F,  $^{20,22}$Ne,  $^{23}$Na,  $^{24,25,26}$Mg, and  $^{27}$Al.
The stars show the observed values for the most iron poor star known today (Frebel et al. 2008). The hatched areas corresponds to the range of values covered by
observed values in CEMP-no stars. The left column corresponds to non-evolved stars (sample of 6 stars), the right column to evolved stars (sample of 17 stars). The observations have been taken
from the SAGA database \citep{Suda2008}.
                  }
        \label{blobZ5}
  \end{figure}

In Fig. \ref{blobZ5} (left panel) we show similar results for the rotating  60 M$_\odot$ at $Z=10^{-5}$ (model E in Table \ref{data}).
For the CNO elements we obtain qualitative similar results as at $Z=10^{-8}$. We can note that the best agreement with observation is also
obtained for shells containing pure H-burning products and both H- and He-burning products.

\subsection{Predictions for Na, Mg and Al abundances in CRUMP stars}

Since the $Z=10^{-5}$  models were computed with account for the Ne-Na and Mg-Al chains, we show on the right panel of  Fig. \ref{blobZ5}  the theoretical ratios
expected for $^{19}$F,  $^{20,22}$Ne,  $^{23}$Na,  $^{24,25,26}$Mg, and  $^{27}$Al.
The Ne-Na and Mg-Al chains are active in H-burning regions.
Their main effect is to transform Ne into Na and Mg into Al. At the beginning, the shells have near solar [X/Fe] ratios (the curve
corresponding to the first shell is the nearly horizontal line just above the [X/Fe]=0 line, see right panel of Fig. \ref{blobZ5}).
The composition is not exactly solar because some surface enrichments have already occurred (see Fig. \ref{fig1wind}) and 
 the initial composition adopted to compute the present models are slightly different
from the \citet{Asplund2005} solar composition which is used to normalise the plotted values.

Then when evolution proceeds, the general trend is the following: [Ne/Fe] decreases and [Na/Fe] increases, [Mg/Fe] decreases and [Al/Fe] increases,
as is expected from the  activity of the Ne-Na and Mg-Al cycle. The sawtooth shape of the abundance distribution pattern is
more pronounced in shells ejected at the end of the evolution than at the beginning.

Some effects might blur this simple picture, as for instance the arrival at the surface of some primary elements.
 During the core H-burning phase there is no possibility to produce primary Na and Al, since
there is no way for the star at this stage for producing Ne and Mg. Thus Na and Al are built up from 
the initial Ne and Mg present in the star. During the core He-burning phase, some Ne, through
$^{16}{\rm O}(\alpha,\gamma)^{20}$Ne in the most massive stars, and Mg, through $\alpha$ captures on $^{22}$Ne 
can be produced at the end of the core He-burning phase. Note that $^{22}$Ne has a strong primary component in rotating models
since this isotope is produced by the transformation of primary nitrogen which has diffused from the
H-burning shell into the He-burning core. 
Since the Ne and Mg are produced at the very end of the core He-burning phase, there is little time left for these
elements to diffuse in the H-burning shell, to be transformed into Na and Mg and to be ejected into the stellar winds. In the present models
no primary components of Ne, Mg, Na and Al are present in the wind material.

Fluorine is built up in regions partially burnt by He-burning (periphery of the convective core
during the core He-burning phase or of the He-burning shell).
It results from transformation of part of
the $^{14}$N. 
Its abundance can be boosted by the production of primary nitrogen in rotating models.
Since it is produced only in a relatively small part of the star, its apparition at the surface
is very sensitive to many parameters of the models. In the model shown in Fig. \ref{blobZ5},
F is slightly depleted (with respect to Fe) in the wind ejecta.


  
When we compare the global shape of the distribution pattern predicted in the shells with the observations, the first striking feature
is the fact that the observations do not show the expected sawtooth pattern between Mg and Al. Magnesium remains more
abundant than Al in HE 1327-2326 and in all the CEMP-no stars for which the abundances of these two elements
have been measured. Thus this seems to be a general feature and not just a peculiar circumstance.

In the frame of the sole activity of the Mg-Al chain, such a pattern might be explained if the material had just
been very partially processed by this chain. In that case 
Mg can be very litttle depleted while Al can still be
significantly increased (since initially Mg/Al is about a factor 10, the transformation of 10\% of Mg into Al would already increase
the Al abundance by a factor 2). This may explain the relative abundance of Mg with respect of Al. 
But a difficulty in that case would be to explain 
the absolute values. Since Mg would be little depleted, the high observed [Mg/Fe] would correspond
to nearly the initial [Mg/Fe] ratio and this would indicate that at [Fe/H] equal to -5.96, the relative number of Mg
atoms with respect to iron is about 100 times greater than the similar ratio in the Sun. 
Looking at the observations of \citet{Cayrel2004} this
does not appear to be the case.   In normal halo stars, values of [Mg/Fe] around 0.2 are found down to a [Fe/H] equal to -4.2. 

Another possibility would be that some Mg and Al produced in He-burning regions are ejected either in the winds or by the supernova.
This is quite a reasonable solution, especially since a very tiny amount of core material would be sufficient. This can be seen
on Fig. \ref{blob60SN} where the chemical composition of material composed of the stellar winds plus supernova ejecta 
is shown for different values of the mass cut (see Sect. 4 for more details).  We see that ejecting at the time
of the supernova only the outermost 7 M$_\odot$  stellar layers  would allow to obtain the correct Mg/Al ratio
(see curve with a mass cut = 30 M$_\odot$ on Fig \ref{blob60SN}). In that case, some dilution is needed for obtaining
the correct enrichment level. This will be discussed in Sect. 4.

   \begin{figure}[t]
  \centering
  \includegraphics[width=9cm, height=9cm]{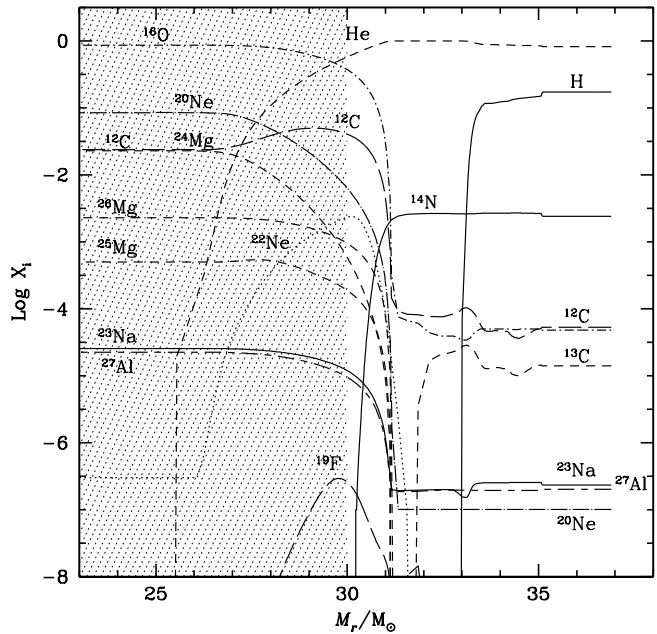}
     \caption{Abundance profiles for various elements  (in mass fraction) as a function of the Lagrangian mass in the outer layers
     of a 60 M$_\odot$ model at $Z=10^{-5}$ with $\upsilon_{\rm ini}=800$ km s$^{-1}$ at the end of the core He-burning phase  (Model E in Table \ref{data}). The grey area represents the part of the star which would remain locked into
     the remnant when the mass cut is equal to 30 M$_\odot$. 
                  }
        \label{chem5}
  \end{figure}
 \begin{figure}[t]
  \centering
  \includegraphics[width=9cm, height=9cm]{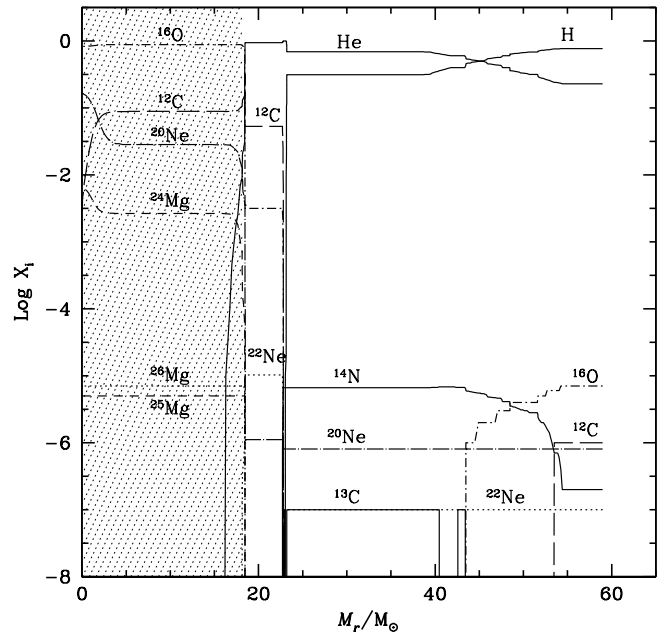}
     \caption{Same as Fig \ref{chem5} for a non rotating 60 M$_\odot$ model at $Z=10^{-5}$. Models computed by \citet{paperVIII2002}. The grey area represents the part of the star which would remain locked into
     the remnant when the mass cut is equal to 18.2 M$_\odot$.
                  }
        \label{chem5N}
  \end{figure}

There is still one striking point to underline when looking at the right panels of Fig. \ref{blobZ5}: in most CEMP-no stars, aluminium
is depleted with respect to iron. This is in contrast with the peculiar case of HE 1327-2326 and does appear difficult to conciliate
with what we know about the nucleosynthesis of aluminium: whatever chain of reactions intervene, aluminium is produced not destroyed! 
At that moment we just mention that Al abundances are determined from resonance lines which are very sensitive to NLTE effects
\citep{Cayrel2004}. The abundances obtained from NLTE models might be 0.8 dex above the LTE abundances
\citep{Andrievsky2008}! In view of this, it might be premature to discuss in too many details solutions of that problem.
We briefly come back to it in the next section.

\section{The spinstar wind and supernova model}
  
       \begin{figure}
  \centering
  \includegraphics[width=9cm, height=9cm]{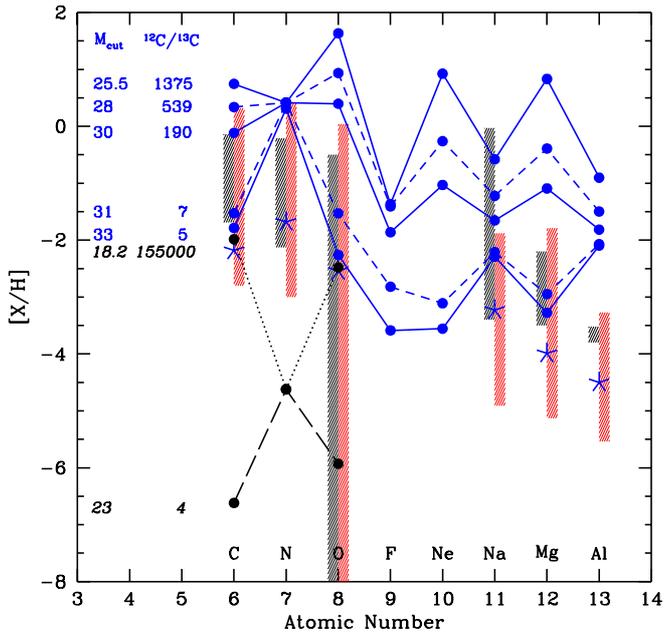}
     \caption{Composition of the mixture composed of wind material of  60 M$_\odot$ models at $Z=10^{-5}$  and of layers ejected by the supernova.  The upper curves (continuous and short dashed lines) correspond to the rotating model
     with $\upsilon_{\rm ini}=800$ km s$^{-1}$ (Model E in Table \ref{data}). The lower curves (dotted lines and long dashed curves) correspond
     to the non-rotating models \citep[models from][]{paperVIII2002}. In those last cases, the curves were shifted downwards
     by 2.7 dex  for clarity.
     Various values of the mass-cut are considered. The
     $^{12}$C/$^{13}$C number ratios in the mixture  are also indicated for each case.
                 }
        \label{blob60SN}
  \end{figure}

In Fig. \ref{chem5}, the chemical structure of our model E in Table \ref{data} is shown. We shall make the hypothesis
that the outer layers of this model already have, at the stage represented on this figure (end of core He-burning), the chemical
composition they would have at the presupernova stage. This is only approximatively correct, but for the elements we are
interested in here, it is a reasonable hypothesis. 

For deciding what fraction of the mass is ejected at the time of the supernova explosion, one needs to know the mass
cut between the ejected shells and the layers which finally will end into the compact stellar remnant. Here we 
consider the value of this mass cut as a free parameter. In case this mass cut is chosen equal to the
presupernova mass, then the supernova does not contribute to the enrichment of
the ISM and the star will contribute only through its winds. For the case when the mass cut is inferior
to the presupernova mass, the mass ejected by the supernova is added to the mass ejected by the winds
during the whole previous lifetime and this constitutes the material composed of wind material and of supernova ejecta.

The composition of the wind+supernova material  is shown in Fig. \ref{blob60SN} for various values of the mass cut.
Down to a mass cut of about 31 M$_\odot$, the supernova ejecta would not change a lot the composition with
respect to the composition of the wind material only. Typically, as explained above, one sees the clear signature
of the activity of the Ne-Na and Mg-Al chains. When He-burning regions are ejected, typically when the mass cut
is decreased from 31 to 30 M$_\odot$, a big change occurs. We see that carbon and oxygen increase a lot
while
the sawtooth pattern typical of the Ne-Na and Mg-Al chains is reversed.  
One notes also that the $^{12}$C/$^{13}$C number ratio increases by orders of magnitude. 
A mass cut as low as 28 M$_\odot$ and beyond would produce O-rich stars.

From Fig. \ref{blob60SN}, we can therefore deduce that to explain nitrogen abundances
superior to those of carbon and oxygen as
observed
in HE 1327-2326 for instance, the mass cut should be high, superior or equal to 30 M$_\odot$
in the case of the model plotted in Fig. \ref{blob60SN} and therefore the
supernova contribution should be small. We see that
adding deeper layers (lowering the mass cut below 30 M$_\odot$) would rapidly lower the N/C and N/O ratios
and thus prevent ratios higher than 1 as observed in HE 1327-2326.
This means that most of the supernova ejecta fall back in the black hole and little amount of the mass of the He-core
is ejected (so not much C and O are finally ejected). This  would be  in line with the model proposed
by \citet{Umeda2003}. Let us note that in that model, in addition to important fall back,
mixing is supposed to occur in a fraction of the He-core at the time of the explosion.  This
mixing process, which was found to take place in SN1987A and various explosion models (see the above reference), allows
some iron, present in the deep layers to be dredged-up in the ejected material. 
The mixing and fallback model also mimics the effect of asymmetric explosions in which matter from deep layers are ejected along the poles while most of the matter falls back on the remnant at other lattitudes.
Such models might be interesting to explain possibly negative [Al/Fe] ratios while still maintaining CNO overabundances compatible with
those observed in CRUMPS.

It is interesting to compare these results with those which would be obtained from a non-rotating model.
The chemical structure of such a model is shown in Fig. \ref{chem5N}\footnote{Since that model was computed
without the account for the Ne-Na and Mg-Al chains, these elements are not shown on the figure.}.  The differences
are striking. The most important ones are the following: the rotating model has lost more than 23 M$_\odot$ by stellar winds, while the non-rotating model has lost nearly no mass.
The envelope of the rotating model is enriched in primary $^{12}$C,  $^{13}$C, $^{14}$N and $^{16}$O.
The envelope of the non-rotating model presents the usual marks of CNO processing, without primary production
of any of the CNO isotopes.
The He-burning shell of the rotating model  is significantly enriched in
primary $^{19}$F  and $^{22}$Ne, and the CO core has a mass of 30 M$_\odot$.
In the non-rotating
model, there is no significant primary $^{19}$F  and $^{22}$Ne and the CO core has a mass of 18.2 M$_\odot$.

The chemical composition of the material composed (mainly) of supernova ejecta from the non-rotating model
is shown in Fig. \ref{blob60SN} for two values of the mass cut (note that the curves for the non-rotating have been shifted downwards by 2.7 for clarity, see caption). When the mass cut is chosen so that only CNO processed material is ejected (mass cut=23 M$_\odot$), then we obtain nitrogen enhanced with respect to carbon and oxygen. The absolute values of carbon and oxygen are very low, of the order of -4 for [C/H] and of -3.5 for [O/H], much lower than the values observed in CRUMP stars.
The abundance of nitrogen is also low, even accounting for the fact that the value plotted in Fig. \ref{blob60SN}
has been shifted by 2.7 dex. The model plotted here has an initial metallicity equal to [Fe/H] $\sim$ -3.3.
Would we have taken a [Fe/H] $\sim$ -6, the abundance of nitrogen would be 2.7 dex lower  (i.e. would 
be at the position plotted in Fig. \ref{blob60SN}).
When the mass cut is 23 M$_\odot$,
the $^{12}$C/$^{13}$C number ratio is quite low in the non-rotating model ejecta. It is typical of the CNO cycle\footnote{However, since here carbon and oxygen abundances depend
on the initial abundances, any mixing with interstellar material would rapidly increase the $^{12}$C/$^{13}$C number ratio up to its value
in the interstellar medium.}.

When the mass cut is chosen to be equal to 18.2 M$_\odot$, nitrogen is not significantly changed, while carbon and
oxygen are increased by many orders of magnitudes.
The $^{12}$C/$^{13}$C number ratio also reaches much higher ratios than in the material ejected
by the rotating model.

\begin{figure}
  \centering
  \includegraphics[width=9cm, height=9cm]{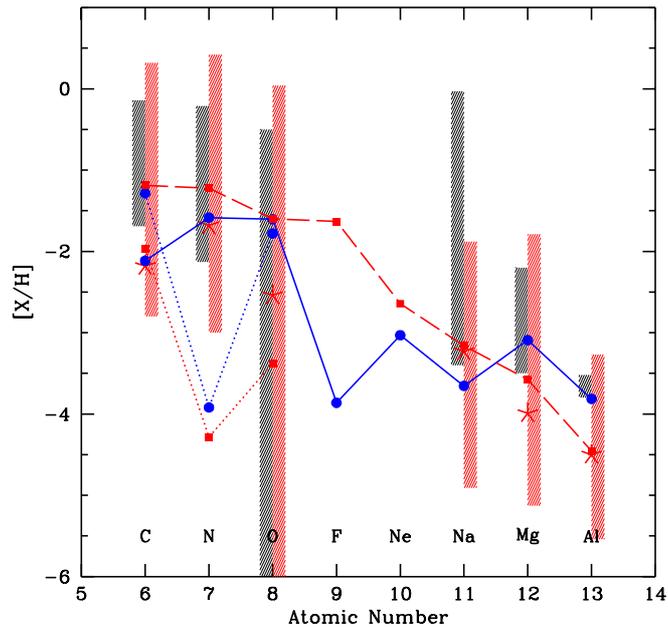}
     \caption{The continuous curve with filled circles shows the
    composition of the mixture composed of material ejected by the rotating 60 M$_\odot$ models at $Z=10^{-5}$
    (wind and supernova, model E in Table \ref{data}) and of interstellar material.
 A dilution factor of 100 has been used (here the dilution factor is
     the mass of interstellar medium divided by the mass ejected by the 60 M$_\odot$). A mass cut equal to 30 M$_\odot$
     was chosen. The dotted curve with filled circles corresponds to the non rotating 60 M$_\odot$ stellar model with a
     mass cut of 18.2 M$_\odot$. The same dilution factor was applied. The dashed line with squares shows the
     composition obtained by mixing the outer envelope (mass above the CO core) at the E-AGB phase of our rotating 7 M$_\odot$ stellar model
     with interstellar medium (dilution factor equal to 100). The dotted curve with squares shows the composition obtained
     in the same manner (mass above the CO core and interstellar medium) using our analog non-rotating 7 M$_\odot$ stellar model.
       }
        \label{blobZ5S}
  \end{figure}

In Fig. \ref{blobZ5S}, we show the composition obtained by the wind and supernova models accounting for
a dilution factor of about 100 with pristine interstellar medium. We obtain a reasonable agreement using our rotating model.
In that case, due to the large dilution factor, the star would not be He-rich. Its $^{12}$C/$^{13}$C ratio
would reach a high value of the order of 190. Thus such a model would have difficulties in explaining
non-evolved CEMP stars showing low $^{12}$C/$^{13}$C ratios.
Such a model would also require an efficient depletion process for lithium.

The non-rotating model would predict C-rich stars, but without strong enhancement of nitrogen (only secondary
nitrogen). Such a model would also predict an extremely high value for the $^{12}$C/$^{13}$C ratio. With dilution, D=100,
the ratio would be about 1550.
 
\section{The spinstar AGB model}

As will be explained in Sect. 6.3, AGB stellar models are invoked as a possible cause for the
CEMP stars. In the models published until now in the literature \citep[see e.g.][]{Suda2004}, the CEMP star has acquired its characteristic surface composition
because it has accreted some AGB wind material at its surface. The model we present here is different
in at least two respects. First we shall suppose that the AGB star which provides the peculiar chemical composition
material originates from a star which was a fast rotator during the MS phase. Second, we suppose that
the whole CEMP star or at least a great part of it is made from a mixture of AGB envelope and of pristine interstellar medium, not only
 its 0.01 M$_\odot$
exterior layers as assumed by \citet{Suda2004}. The situation
in the present model is similar to that sketched in Fig. \ref{schema}, except that, in panels A to C, instead of having a massive star, we consider here the case of an intermediate mass star (panels D1 and D2 of course do not apply in the case of an intermediate mass star).

   \begin{figure}
  \centering
  \includegraphics[width=9cm, height=9cm]{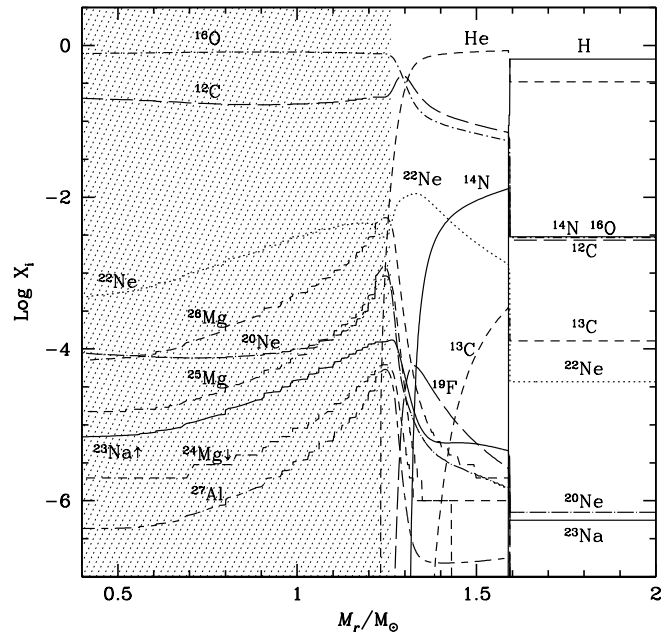}
     \caption{Variation of the abundances of various elements  (in mass fraction) as a function of the Lagrangian mass in the outer layers
     of a 7 M$_\odot$ model at $Z=10^{-5}$ with $\upsilon_{\rm ini}=450$ km s$^{-1}$ at the end of the core He-burning phase  (Model F in Table \ref{data}). The grey area corresponds to the CO-core.
                  }
        \label{chemagb}
  \end{figure}
  
     \begin{figure}
  \centering
  \includegraphics[width=9cm, height=9cm]{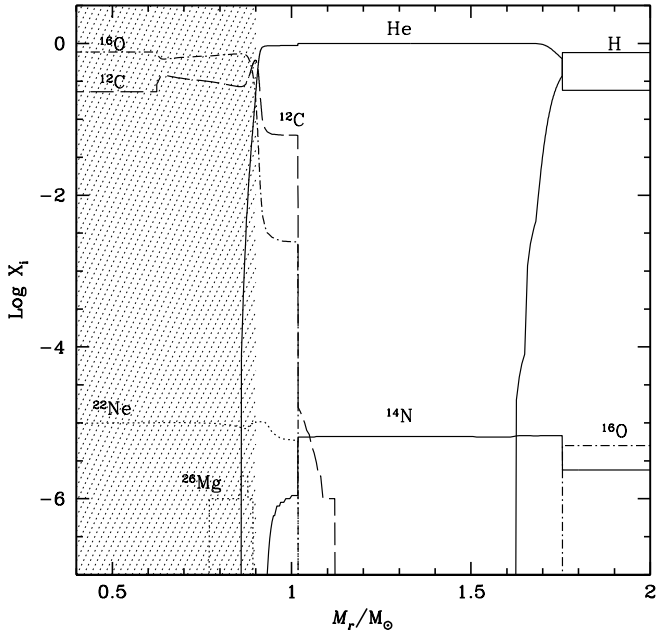}
     \caption{Same as Fig \ref{chemagb} for a non rotating 7 M$_\odot$ model at $Z=10^{-5}$. Model computed by \citet{paperVIII2002}.  The grey area corresponds to the CO-core.
                  }
        \label{chemagbN}
  \end{figure}
  
Before making the comparisons with observed abundances at the surface of CEMP-no stars, we want to make two remarks.
First our 7 M$_\odot$ stellar mass model is at the early-AGB phase. During the subsequent phase, the thermal pulse AGB phase, the
chemical structure of the envelope will change and strong mass loss will occur.
Thus the present models should be pursued in the TP-AGB phase before any more
quantitative conclusions can be reached. At the moment, the results presented here may
however be relevant in the frame of the close binary scenario.
If our intermediate mass star belongs to a close binary system,  some mass transfer may occur when the primary inflates to join the early asymptotic giant branch; in
that case some mass can be accreted by the secondary and/or lost by the system as a whole\footnote{Such a scenario can occur in a binary wide enough for the star
to evolve into the E-AGB phase, and tight enough so that mass transfer occurs
during the E-AGB phase.}.

The second remark, is that according to the $Z$ = 10$^{-5}$ models of \citet{paperVIII2002}, computed for initial velocities of 300 km s$^{-1}$, similar
behaviours are obtained at least down to 2 M$_\odot$. Thus, rotational mixing will have an impact, not only on the chemical structure
of the most massive AGB stars, or those rotating fast. It will have an impact for a whole range of initial masses and rotational velocities.

In Figs. \ref{chemagb} and \ref{chemagbN} the chemical structure of our rotating and non-rotating 7 M$_\odot$ models at $Z=10^{-5}$ are shown.
As for the most massive stars, the differences
are striking and are qualitatively the same. The only point which is very different with respect
to the massive star case, is that up to this point, even the rotating intermediate mass star model has lost very little amount of mass
by stellar winds. This comes from the fact that the surface enrichments in the rotating model occurred just before
the evolutionary stage shown in Fig. \ref{chemagb}  and thus the enhanced mass loss had no time to peel off the star.

We can note that, in the rotating model, in the zone between the He-burning shell (at about 1.3 M$_\odot$) and the H-burning one
(at 1.6 M$_\odot$), the levels of primary $^{13}$C, $^{14}$N and $^{22}$Ne are higher than in the rotating 60 M$_\odot$
model. This comes from the fact that the 7 M$_\odot$ evolves rapidly to the red part of the HR diagram at the end
of the MS phase, as a result of the rapid core contraction and envelope expansion. Thus a strong $\Omega$-gradient is present at the border of the He-burning core during most of the
He-burning phase, making possible an efficient diffusion of He-burning products into the H-burning shell and of
H-burning products into the He-burning core.

On Fig. \ref{blobZ5S} we show the composition of a star made of the envelopes of such AGB stars mixed with
interstellar material (dilution factor equal to 100). We see that we obtain a reasonable fit if we consider our
rotating model. Such a star would not be He-rich, and would present a $^{12}$C/$^{13}$C ratio equal to 104.
Our non-rotating model cannot account for the high nitrogen abundance. 
Moreover, the $^{12}$C/$^{13}$C ratio would be of the order of 8000 (after dilution).

The spinstar AGB model would however have some difficulties explaining the lower than solar [Al/Fe] ratios (if real!). Such
stars do not eject any Fe, while they inject some Al, thus they would predict positive [Al/Fe] values as is observed in HE 1327-2326!

\section{Alternative models}

In this section we shall briefly discuss alternative models for explaining CRUMPS.
Alternative models invoke either AGB stars in binary systems or supernovae.  \citet{Venn2008} on the other hand suggest that the abundance pattern observed today at the surface
of CRUMPS does not reflect the initial abundances of the cloud from which the star was born but
results from processes occurring in or around the star. Let us begin by examining this hypothesis.

\subsection{Are CRUMPS produced by in situ processes?}


\citet{Venn2008} suggest that CRUMPS are not actually very metal poor stars, but may have their photosphere deprived in heavy elements due to a separation of gas and dust beyond the stellar surface, followed by the accretion of dust-depleted gas. Such process is invoked to explain Lambda Bootis stars, which are MS stars with spectral type from A to mid-F, found all along the MS phase. They are expected to have a solar-like composition but show significant metal deficiencies. The authors propose some tests of their hypothesis, for instance to look at the abundances of sulfur and zinc. Dust formation does not deplete so much these two elements and thus
in the frame of this scenario they should not show the same depletion as the rest of metals. 
Unfortunately no such observations are available yet. 

While this explanation is indeed quite interesting, it can probably not be invoked to explain 
all the observed cases. Our main argument is the following: if the dust-gas separation process may produce different elemental ratios, it should not change
isotopic ratios!  
Such a scenario have difficulties to explain why for instance in G77-6 \citep{Plez2005}, which is a non-evolved star,
values as low as 5 of the $^{12}$C/$^{13}$C ratio have been measured. This value is well below the
$^{12}$C/$^{13}$C ratios measured at the surface of stars belonging to the ``unmixed'' sample of red giants metal poor stars observed by \citet{Spite2006} and thus results from a nuclear process
having occurred in stars of previous generations and not from an in-situ fractionation process. 

  \subsection{Non-rotating supernovae models}
  
CRUMPS might form from material made of a mixture of supernova ejecta (originating from non-rotating progenitors) and of interstellar material.
A question which immediately arises when suggesting this origin for CRUMPS is why are CRUMPS produced at very low metallicity? From physical considerations, one expects that supernovae originating from zero metallicity or very metal poor progenitors present different characteristics from supernovae coming from solar metallicity progenitors. At zero or very low metallicity,  progenitors are more compact. This favors higher amount of fallback \citep{Chevalier1989}. The more compact structure of the presupernova at very low metallicity reduces the extent of the region where Rayleigh-Taylor instabilities grow and where the mixing of the ejecta can occur. A consequence of this is that the yields of such supernova explosion are very sensitive to the amount of fallback.
Thus it may be that the ejecta at very low metallicity present characteristics very different from those expected at standard metallicities.
 
Many models invoking supernovae have been proposed to explain the peculiar abundance pattern of CEMP stars \citep{Umeda2003, Limongi2003, Tominaga2007b,Joggerst2009}.
For instance, the zero-metallicity models of \citet{Joggerst2009} show great overproductions in carbon and oxygen relative to iron, a trend corresponding to what  is observed in CRUMPS. However such models have difficulties to reproduce the strong nitrogen overabundances, quite in line with what
we have obtained with our non-rotating models (see above).
As indicated by these authors, accounting for the effects of rotation might help in resolving this difficulty. This does indeed appear to be the case!

Another difficulty which may come out from such models is that they need very high dilution factor with pristine interstellar medium in order to reproduce
the observed levels of the overabundances. As we have seen this is contrary to
the observed very low Li abundances unless depletion mechanisms much stronger than those usually accounted for in present day models is supposed
\footnote{Looking at Fig. 3 of \citet{Tominaga2007b} it might seem that
a star produced from supernova ejecta with high dilution with ISM would produce Li depleted stars. This is an artefact due to the fact that what has been plotted are the ratios
of the abundances of these elements with respect to iron in the ejecta alone. For heavy elements this is equal to the ratio of the elements after dilution since the primordial matter contains no heavy elements. However
for H, He, D and Li this is not the case. We checked that this interpretation is correct (Tominaga, private communication).}.
Even in case of very strong Li depletion mechanism, such a model cannot account for those CEMPS stars with low $^{12}$C/$^{13}$C number ratios (see also Sect. 7).

Supernovae implies large dilution factors, therefore such models do not predict He-rich CRUMPS! 
Detection of He enhancement in CRUMPS would imply small dilution factors and thus eliminate a strong contribution by supernovae ejecta (at least in He-rich CRUMP stars).

\subsection{AGB stars in binary systems}

Many authors argue that the peculiar surface abundances of CEMP stars have been acquired 
in a mass transfer episode. The most commonly accepted scenario follows
the following general line: when the primary has evolved into the AGB phase, part of its wind material
is accreted by the secondary (today the CEMP star) giving
it the peculiar surface abundances that we can observe today.

The following facts have been invoked as supporting the AGB binary scenario:
\begin{itemize}
\item The presence of {\it s}-process elements in CEMP-{\it s} stars is interpreted as a mark of the AGB nuclesosynthesis.
\item The very low iron abundances imply a nucleosynthetic source which does
lead to very weak or no iron enrichment. AGB stars do not produce iron.
\item  Binarity seems indeed to be observed
(through radial velocity variations) for most of the CEMP-{\it s} \citep{Lucatello2005}.
\item There seems to be a continuity of properties between the C-enhanced stars and the CH-stars \citep{Masseron2009}.
\item A similar argument seems also to hold between  the CEMP-no and the CEMP-{\it s} with
a weak overabundance in {\it s}-process elements  \citep{Masseron2009}.
\end{itemize}

In the scenario  proposed by \citet{Suda2004} \citep[see also][]{Komiya2007},
the progenitor of the CEMP star, was the less massive component in the binary 
with a mass of the order of 0.8 M$_\odot$, 
while the massive component had initially a mass in the range of 1.2 - 3 M$_\odot$. 
The primary experiences a helium-flash driven deep mixing as found by \citet{Fujimoto1990}. In low mass models this occurs when the star is at the tip of the RGB, in
intermediate mass models,  this occurs at the beginning of the AGB phase.
As a result, its surface composition changes. During the thermal-pulse AGB phase, additional material is dredged-up to the surface and the primary star emits a wind. Part of the wind is accreted by the secondary and produces the abundance pattern now observed at the surface of CEMP stars. These authors estimate that in the case of HE0107-5240 \citep{Christlieb2004},
the amount of mass accreted by the secondary is of the order of about 0.01 M$_\odot$ (this estimate
is obtained by requiring that, when the secondary becomes a red giant, the abundances at its surface
match those of HE0107-5240). In this scenario the presence of iron-group elements can be attributed
to the fact that either the binary system formed from a cloud having such an abundance in iron-group elements or by accretion after birth from primordial material of matter enriched in iron-group elements ejected by one or more first generation supernovae. Since the {\it s}-process would not be the same in these cases, values of the ratio [Pb/Fe] may be used to disentangle these two
secenarios \citep{Suda2004}. At the moment current observations do not allow to choose between them.

This scenario presents a reasonable way to explain the CEMP-{\it s} stars. Among its advantages, beside
explaining the peculiar surface abundances of CEMP stars, it may provide an explanation why the C-rich stars are more frequent at low metallicity. Let us recall that about 20\% of stars with [Fe/H]$\le -2.0$ exhibit enhancement of the surface carbon abundance \citep{Lucatello2006}, while only 1\% of Pop II stars are characterized by enhanced CH bands in the spectra \citep{Luck1991}.
The higher frequency of C-rich stars at very low metallicity might be due to the fact that
the helium-flash driven deep mixing phenomenon does appear as a distinct feature of the evolution of extremely metal poor stars of low and intermediate mass. Also a companion has been detected around 68\% of CEMP-s stars \citep{Lucatello2005}. 

In that scenario the peculiar abundance results from the accretion of a very small amount of material,
0.01 M$_\odot$ for matching HE0107-5240 which is a red giant \citet{Suda2004}\footnote{Note that HE0107-5240 is 
not a CEMP-s star. Despite that, the case of this star is discussed in detail in the paper  by \citet{Suda2004}  in the frame of the AGB binary scenario.}! Still smaller amounts of material
would be necessary to reproduce the abundances observed at the surface  of non-evolved stars presenting similar surface enrichments. 
This means that obviously such a scenario would not predict He-rich stars.

While this scenario is quite successful in many respects, it has also some drawbacks. For instance,
according to  \citet{Stancliffe2008}, (non-rotating) AGB models do not predict substantial enhancements 
of both carbon and nitrogen at the same time (except in a very narrow mass range).
The spinstar AGB model presented previously would provide a way to overcome that difficulty.

\section{Possible origin of CRUMP stars with $^{12}$C/$^{13}$C below 30}

In this section we want to specifically address the origin of CRUMP stars showing a low  
 $^{12}$C/$^{13}$C isotopic ratio, typically below 30.  Interestingly in the SAGA databasis,
 except for two stars, all the other CEMP-no stars (7 stars) present $^{12}$C/$^{13}$C
 below 25. The two exceptions have upper limits in the range of 40-50.

The  $^{12}$C/$^{13}$C isotopic ratio has four properties which makes it a wonderful probe of various
nucleosynthetic environments:
\begin{enumerate}
\item In a low mass main-sequence or subgiant star as HE 1327-2326, there is no known process which
may modify this number ratio at the surface. Thus this quantity has the advantage over lithium that
its present day value at the surface of non-evolved stars is equal to the initial value of this ratio.
\item The $^{12}$C/$^{13}$C number ratio presents great variations depending on the source material
considered. 
\item Isotopic ratios depend on the nuclear paths, they are not sensitive to fractionation effects.
\item Finally from a nucleosynthetic point of view, low $^{12}$C/$^{13}$C can only be obtained
when little amount of pure He-burning material is ejected.
\end{enumerate}

The number ratio in the CRUMP star can be expressed as
$${n(^{12}{\rm C}) \over n(^{13}{\rm C})}= {13 \over 12} {X(^{12}{\rm C})_{\rm source}+X(^{12}{\rm C})_{\rm 0} D \over X(^{13}{\rm C})_{\rm source}+X(^{13}{\rm C})_{\rm 0} D},$$
where $n$  is the number fraction of the element considered and $X$ its mass fraction in the source material (subscript ${\rm source}$) or
in the interstellar medium (subscript ${\rm 0}$). 
Thus we need to know the values of $X(^{12}{\rm C})_{\rm 0}$ and of $X(^{13}{\rm C})_{\rm 0}$.
Unfortunately, these values are
not known for the very low [Fe/H] corresponding to CRUMPS. According to \citet{Spite2005},
values of [C/Fe] around solar values are found in giants with [Fe/H] between -4.2 and -2.1.
\citet{Spite2006} give the number ratio $^{12}$C/$^{13}$C obtained in giants at low [Fe/H] (down to about -3.5), that have undergone
the first dredge-up but are still below the luminosity bump. Values between 20 and 30 are obtained.
These values are lower than the initial value since they were
decreased by the effect of the first dredge-up. According to \citet{Chiappini2008}, accounting for the effect
of the first dredge-up implies initial values for $^{12}$C/$^{13}$C around at least 50.  

Of course there is no guarantee that these values are the same at
much lower values of [Fe/H]. If one uses the chemical evolution model of  \citet{Chiappini2008}, we obtain at 
[Fe/H]=-5, values which differ greatly if one considers yields from slow rotating or fast-rotating models:
slow rotating models, {\it i.e.} with angular momentum lower than those of massive stars at solar metallicity, would yield
very high values around 31 000!  Fast rotating models,  {\it i.e.} with angular momentum content of similar magnitude as that of
massive solar metallicity stars, predict ratios equal to 30. These last models are to be preferred since they also
well reproduce the [N/O] and [C/O] observed trends with [Fe/H] and [O/H] in the halo \citep{Chiappini2006}.
Whatever the correct answer is, we can say that probably the initial $^{12}$C/$^{13}$C ratio at very low metallicity
is greater than 30.

The effect of dilution is very different depending on the primary or secondary nature of the nucleosynthetic
path leading to the synthesis of the two carbon isotopes in the source material. In case both isotopes have a strong primary
component as is the case in spinstar models\footnote{Whatever the spinstar model considered: AGB envelope, wind material or supernova ejecta.},
then the $^{12}$C/$^{13}$C ratio will not be much affected by dilution. Typically at $Z=10^{-8}$ the initial abundances of $^{12}$C and $^{13}$C, that we have
used to compute the models are 7.55 10$^{-10}$ and 0.10 10$^{-10}$ respectively, while the abundances in the winds of the 85 M$_\odot$ model at $Z=10^{-8}$ are 0.097  ($^{12}$C) and 0.009 ($^{13}$C)
(see Table \ref{data}).  Thus, for any  dilution factors smaller than about 10$^{8}$, the number ratio, equal to 11, is given
by ${13 \over 12} {X(^{12}{\rm C})_{\rm source} \over X(^{13}{\rm C})_{\rm source}}$ and is independent of the dilution factor. 

In the spinstar models, the only way to obtain values for the $^{12}$C/$^{13}$C ratio below 30 is to consider only wind
material of massive stars with no or little amount of pure products of He-burning. 
Note that models which would lose this envelope not by winds but at the end of the evolution when they
explode as supernova, would eject material with nearly identical compositions as the wind models presented here.

Would rotating AGB models be able to produce low $^{12}$C/$^{13}$C number ratios?
If we consider the mass fractions of the two carbon isotopes in the envelope of our 7 M$_\odot$ model, we have 8.8 10$^{-3}$ for $^{12}$C and
0.12 10$^{-3}$ for $^{13}$C. The initial abundances at Z=10$^{-5}$ are respectively 7.55 10$^{-7}$  ($^{12}$C) and 0.10 10$^{-7}$  ($^{13}$C) thus at least three 
orders of magnitude lower than the abundances in the envelope at the early AGB phase.  If the dilution factor is well below 10000 then one would obtain a  $^{12}$C/$^{13}$C number ratio equal to 79 and if the dilution factor
is greater than 10000, one would obtain a  $^{12}$C/$^{13}$C number ratio equal to that of the interstellar medium, between 30 and 31000 (see the discussion above)!
Therefore, it does appear
that our early-AGB (E-AGB) model would not predict very low $^{12}$C/$^{13}$C number ratio, but this is only one model at a peculiar stage. At the moment
we cannot exclude that AGB material might also yield very low $^{12}$C/$^{13}$C material.

For non-spinstar models, we discuss here only the case of the supernova models, for which the published
data allow an easy calculation of the expected $^{12}$C/$^{13}$C ratio in the CRUMP star.
Let us take the yields of the 40 M$_\odot$ Pop. III supernovae
computed by \citet{Tominaga2007b}.  We consider their model 40B in their Table 2. Using their mass cut of 10.7 M$_\odot$, we obtain mass fractions
of $^{12}$C and $^{13}$C in the supernova ejecta equal to 0.012 and 2.7 10$^{-9}$. Mixed with primordial material, one would obtain a $^{12}$C/$^{13}$C number ratio equal to
4.4 10$^6$ independent of the dilution factor. Similar values would be obtained from other models by  \citet{Tominaga2007b}. Note that this is
not a conclusion linked to the peculiar supernova model cited above. The supernova ejecta will be richer than the winds in $^{12}$C and, at least, if they
are computed from non-rotating models, would be relatively poor in $^{13}$C. Thus this means that the $^{12}$C/$^{13}$C number ratio will be very high!

From the above considerations, it appears that a non-evolved CEMP star with a $^{12}$C/$^{13}$C number ratio inferior to about 30 cannot be made 
from (non-spinstar) Pop III supernova ejecta whatever the dilution factor. It can be made from the winds of spinstars even highly diluted with interstellar material.

\section{Conclusion and perspectives}

\subsection{Some general considerations}

\begin{table*}
\caption{Theoretical interpretations of observed features and consequences for helium in CRUMP stars.
}
\scriptsize{
	\centering
		\begin{tabular}{clllc}
		\hline
		\hline 
N & Observed feature  & Hypothesis  &  Theoretical interpretation   & He-rich \\
	               \hline \\
1 &$\epsilon$(Li) $<$ Spite     & no in-situ depl.           & $D < \sim 0.5$                                                                  &  YES                                      \\
    &                                                     &                                       &                                                                                             &                                           \\         
2 &$\epsilon$(Li) $<$ Spite     & strong in-situ depl.    & $D >$ 1                                                                            & YES   for $D< \sim 4$          \\
   &                                                      &                                        &                                                                                           &                                            \\   
3 &$\epsilon$(Li)  $\sim$ Spite &                                     & $D >$ 1, in situ Li- depletion                                         & YES   for $D< \sim 4$        \\
  &                                                       &                                       &                                                                                           &                                            \\  
		   &     & & &  \\                                                                                      
4 & $^{12}$C/	$^{13}$C $<$ ISM                                   &                                                                                            & H- and He-burn. and no He-only                          &  ?                                          \\
5 & $^{12}$C/	$^{13}$C $>$ ISM                                   &                                                                                            & H- and He-burn. plus He-only, $D \gg 1$           &  NO                                      \\
		   &     & & &  \\
6 & [N/H] $>$ 1.4+[Fe/H]                                                    &  0 $\le$ ISM [CNO/Fe]$\le$0.6                                      & source material is primary N rich                  & ?                                            \\
		   &     & & &  \\   
7 & [N/C] and [N/O] $>0$                                                         &                                                                                              & H- and He-burn.  and no He-only                        & ?                                          \\     
8  & [N/C] and [N/O] $<0$                                                        &                                                                                              & H- and He-burn. plus He-only, $D \gg 1$           &  NO                                    \\
		   &     & & &  \\
9& [Na/Ne] and [Al/Mg] $>0$                                                 &                                                                                              & H- and He-burn. and no He-only, small $D$               & YES                            \\ 
10& [Na/Ne] and [Al/Mg] $<0$                                              &                                                                                               &  H- and He-burn. plus He-only, $D  \gg 1$    &  NO                                \\
		   &     & & &   \\
11& [Al/Fe] $>0$                                                              &  ISM [Mg/Fe] $\sim$ 0                                                         &  H- and He-burn.                                                     &  ?                                    \\                
12&[Al/Fe] $<0$                                                                &                                                                                                & [Al/Fe] $< 0$ in  M$_{\rm source}$, $D \gg 1$   & NO                                \\
\hline   
		\end{tabular}
		\label{interp}
		}
\end{table*}

In this subsection, we want to collect some general conclusions which are not
dependent on a peculiar model for the source of material which is responsible 
for the abundances observed in CEMP stars. 

In Table \ref{interp}, we enumerate 12
observed features (see column 2). Each of these observed feature is given a number N (for further reference). It is
indicated in column 1. Column 4 indicates what can be concluded from the observed
feature and the hypothesis given in column 3. 
The CEMP stars are supposed to be composed
of some amount of source material, $M_{\rm source}$ (Li-free) and of interstellar medium material (with WMAP Li).
In column 3, depletion refers to
Li and the symbol ``ISM[X/Fe]''  indicates the value of the ratio in the interstellar medium at the time of
CRUMP/CEMP star formation.
In column 4, $D$ is the dilution factor, ``H and He burn.'' means that the source material from which
the CRUMP/CEMP star is made of  is
rich in both H- and He-burning products, ``no He-only'' means that this material does not contain
(or very little amounts of) matter only processed by He burning, while ``+ He-only''  means that
the source material contains a significant amount of pure He-burning products. Column 5 indicates whether the CRUMP/CEMP
star is He-rich (Yes) or not (NO). A question mark indicates that both cases are possible. 
The following general comments are worth making :

\begin{enumerate}
\item In case Li is not depleted by the CEMP star itself, then any low observed Li abundance implies
a very small dilution factor (feature 1).
\item In case Li is observed to be low and Li is depleted in the CEMP star, the dilution factor cannot be constrained from 
the present-day Li surface abundance (features 2).
\item In case Li is observed to have a value along the Spite plateau, then  the dilution factor cannot be constrained from 
the present-day Li surface abundance and the depletion if any, would be similar to those in normal halo field stars (feature 3)
\item Small $^{12}$C/$^{13}$C ratios (below 30) indicate that the source material has been strongly processed  by the CNO
cycle. Let us note that the high abundance of carbon (which is in the form of $^{12}$C) implies large primary $^{13}$C
production. This can occur only if some mixing has occurred in the source between the He- and the H-burning region (feature 4).
\item Large $^{12}$C/$^{13}$C ratios (above 30) indicate either important dilution factor and/or 
that the source material is dominated by He-burning products (feature 5).
\item The source material is rich in primary nitrogen for CRUMP/CEMP stars showing feature 6.
Let us recall that a secondary process inside the star
can at most produce [N/H] value about 1.4 dex above [Fe/H]. This is true of course as long as the initial
abundances of carbon and oxygen are not orders of magnitudes more abundant with respect to iron
than in the Sun. At the moment, looking at the data by \citet{Cayrel2004}, [C/Fe] is solar and [O/Fe] is about
3 times solar in metal poor normal halo stars.
\item The [N/C] and [N/O] ratios are a measure of the importance of CNO processed material 
in the source material. If these ratios are large, this indicates a large proportion of CNO processed material
(feature 7). Note however that as indicated many times previously, He-burning products are not absent
since they allow the overabundances with respect to iron of carbon and oxygen.
\item If the  [N/C] and [N/O] ratios are small, this favors He-burning processed material (feature 8). 
This is independent of the dilution factors
because all these elements are primary and thus in much greater abundance in the source material
than in the interstellar medium. Only, probably, unrealistic large dilution factor, can change the ratios obtained
in the mixture. Let us note also that one can have [N/C] $<$ 0 and [N/O] $>$ 0 or the inverse in material
ejected with both H- and He-burning material and some pure He-burning material.
\item In a similar way, large [Na/Ne] and [Al/Mg] ratios point towards material having been mainly processed
by the Ne-Na and Mg-Al chains active in hot H-burning regions (feature 9). 
A word of caution however has to be made here. Since
the pattern of the Ne-Na, Mg-Al chains results mainly from a secondary process, dilution can modifies
somewhat the final shape of the abundance pattern. Indeed the elements between Ne and Al have probably
initial abundances of the same order of magnitude and therefore the transformations of for instance 90\% of
Mg into Al will produce contrast between the source material and the interstellar material of at most
one order of magnitude. Such a small contrast can thus be erased by a small dilution factor. 
\item Small  [Na/Ne] and [Al/Mg] ratios are indicative
of material processed in He-burning zones (feature 10).  
\item Feature 11 is compatible with source material processed by the Mg-Al chain.
\item Feature 12 implies that some iron is ejected by the source.
\item  If  He-rich CRUMP/CEMP stars exist, they should be found among
the stars presenting the features 4,6,7,9,11 indicated in Table \ref{interp}. Note that stars
showing these features are not necessarily He-rich.  
\end{enumerate}

\subsection{The main results from our spinstar models}

{\it One of the point which appears from that study is that rotation does play
a key role in explaining the abundance pattern observed in CEMP stars. This is true
whether the source material come from massive star winds, from intermediate mass star winds
or from a supernova}.

The spinstar scenario that we have presented has consequences for
other topical questions related to the chemical evolution of the galactic halo.
First, this scenario is indirectly supported by the fact
that fast rotating stars seem to be needed to reproduce the high N/O plateau
observed in the halo of our Galaxy and also the C/O upturn towards low metallicities \citep{Chiappini2006}.
Hence, an advantage of our scenario is that 
the same kind of objects can be responsible for the properties of both normal
halo stars and the CRUMPS. The normal ones are made up of a well mixed reservoir
which has been enriched by massive rotating stars of different masses and metallicities as proposed by \citet{Hirschi2007}.
The CRUMP stars are formed in the vicinity of one massive (or intermediate) rotating
star and are made up of material taken from the external layers of such stars.

Second, spinstars may also be invoked to explain the chemically peculiar stars
observed in globular clusters\footnote{It is striking to note that chemically peculiar stars found in the field and in the
globular clusters are quite different (rich in both H- and He-burning products in the field and
rich in H-burning products in the clusters). A possible explanation for this fact is proposed in \citet{MeynetIAU268}.}.
At the metallicity  of the globular clusters, fast rotating stars may lose large amounts
of mass when the star reaches the critical limit. This only occurs during the MS phase. The material in the disk
is thus only enriched in H-burning products. This material is released very slowly in the cluster interstellar medium
and can thus be retained by its shallow gravitational potential. It can then be used to form stars bearing the signature
of H-burning processes as explained in \citet{Decressin2007a, Decressin2007b}. When the star evolves further, fast winds occur
which escape from the cluster. The same is true for the supernova ejecta. 

As sketched in our Fig. \ref{schema} we see that some of the CRUMPS stars may receive some
SN ejecta  (see panel D2). This may explain some r-process enrichment. For what concerns
the s-process, we think that rotation may strongly affect the nucleosynthesis of these elements
and that massive stars might be the source of a much broader range of {\it s}-process elements than
presently believed \citep{Pignatari2008}. Also the impact of rotation in the {\it s}-process element
synthesis in intermediate mass star at very low metallicity has also to be investigated in the line
of the work of \citet{Siess2004}.

\begin{table*}
\caption{Spinstar models compatible with sets of observed features given in Table \ref{interp}.
}
	\centering
		\begin{tabular}{llc}
		\hline
		\hline
Observed features  &  Model   & He-rich  \\
	          \hline                                                                                  
  1,4,6,7,9,11  &   spinstar wind model & YES \\
  2,4,6,7,9,11  &   spinstar wind model & ?  \\
  3,4,6,7,9,11  &   spinstar wind model & ?  \\
  2,5,6,8,10,11 & spinstar wind and faint supernovae model or spinstar E-AGB model  & NO  \\
 3,5,6,8,10,11 &  spinstar wind and faint supernovae model or spinstar E-AGB model  & NO  \\
 2,5,6,8,10,12 & spinstar wind and faint supernovae model                                                 & NO  \\
                 \hline	   
		\end{tabular}
		\label{models}
\end{table*}

In Table \ref{models}, we indicate which spinstar model does the best job at reproducing the various
sets of compatible observed features listed in Table \ref{interp}.
Below we draw some conclusions: 
\begin{enumerate}
\item CRUMP stars originating
from the winds of spinstars present strong signs of material processed
by H-burning, {\it i.e.}, high [N/C] and [N/O] ratios,
and small  $^{12}$C/$^{13}$C ratios. This is independent of the
dilution factor. If the dilution factor is small, these stars are
He-rich and Li-poor.
\item The spinstar wind model does appear in the present study as the only model
which can explain CEMP stars with low $^{12}$C/$^{13}$C ratio. It might be that new
rotating AGB star models could also be potential interesting candidates but
this has to be confirmed by further computations. 
\item Some CRUMPS can also be formed of a mixture of both wind
and SN ejecta from massive fast rotating stars.
Supernovae from non-rotating
models cannot reproduce the
abundance pattern observed in CRUMPS.
\item Stars made of the envelope of E-AGB stars resulting from rotating models
present also abundance pattern in qualitative agreement with CEMP-no stars.
Similar non-rotating models do not fit the observed abundances.
\item Both the wind+supernova model and the spinstar AGB model requires
some dilution with pristine interstellar medium to obtain a reasonable fit. This means that
such stars would not be He-rich and would present high $^{12}$C/$^{13}$C ratios.
\end{enumerate}

\subsection{Why are there no analogs of CRUMPS at higher metallicities?}

One may wonder why the scenario leading to the formation of CRUMPS no longer
works in the present universe. 
Two reasons can be invoked:
1) At high metallicity, rotation cannot produce primary nitrogen and thus the peculiar 
simultaneous overabundance of CNO elements.
2) At high metallicity, the chemically enriched winds of massive stars propagate in a medium which
has already been pushed away by the strong and fast O-type star stellar winds. This may prevent
the formation of stars made almost completely of pure chemically enriched ejecta. At very low metallicity the situation
can be very different, the line driven stellar winds are first very weak preventing the star from pushing out
the circumstellar material. Only when the stellar 
surface is enriched in heavy elements,  does the star lose large amount of material by stellar winds.
Therefore, these enriched, slow winds (slow because ejected when the star is a supergiant), 
will have to go through a dense circumstellar medium. This may favor the formation of clumps.
Further detailed hydrodynamical modelings are needed to confirm or reject such scenarios.

\subsection{How to show that some CRUMPS are He-rich?}

Let us end by saying a few words about the possibility to test the hypothesis of CRUMPS  (or maybe also some
CEMP) stars being He-rich. 
Although detection of He will remain problematic in  cool stars, it is not hopeless. 
Among the possibilities let us cite the following ones:
\begin{itemize}
\item Helium lines may be observable in cool stars,
whose strength might be related to the He abundance.
The line He I 10 830 \AA\    was used by \citet{Moehler2000}  mainly for velocity diagnostics. 
It is not a
straightforward line to interpret because it arises from a very excited
metastable level in He I.  However, according to Dupree (private communication)  if stars of
similar $T_{\rm eff}$ are observed,   a relative measure could be done. It might also be possible
to get abundances  but that would  need probably  non LTE modeling\footnote{In the hotter HB stars (T$_{\rm eff}$ $>$10000) other Helium lines become
visible and those have been used to determine abundances \citep[see][]{Moehler2000}.}.
An additional difficulty with this approach comes from the fact that atomic diffusion would
deplete the surface in Helium \citep[see the models by][]{Korn2009}.
\item Asteroseismology may help in revealing the He-abundance of these stars. 
\item To observe an eclipsing binary system whose components are CRUMPS or CEMP stars
and to determine the positions of the two components in the HR diagram as well as
their masses.
The comparison with the stellar models may put in evidence the need to enhance the
He-abundance in order to simultaneously reproduce all the observed features.
\item The presence of a significant population of extremely blue horizontal branch stars in the halo field
may be an indirect hint of He-rich stars, since such stars can be the descendants of He-rich stars.
\end{itemize}

\subsection{What to conclude if He-enhancement is detected? If He abundance is normal?}

If He-enhancement is detected and found to be a feature concerning the whole star and not only its surface, then
 this would definitively imply that the material from which
the CRUMPS are made do contain relatively small amount of interstellar medium.
This would discard models explaining CRUMPS as formed by the accretion of a small amount of mass
at the top of the secondary in a binary system. It would also discard the normal supernovae as
a possible source for the material from which the CRUMPS are made of, because 
the dilution factors which are used to bring the composition of normal supernova ejecta to the level
of the  observed abundances in CRUMPS are much too high to make the star He-rich.
Faint supernovae with large fall back would remain interesting candidates together with
the effect of rotation to explain the high nitrogen enhancements. As discussed in the present work, the wind of spinstars would
also be a favored candidate in that case. Let us stress here that this model is  not incompatible with
the faint supernova model and both may be linked in a same evolutionary history, rotation allowing
to enrich the envelope in H- and He-burning products and to eject it as a wind before the
faint supernova event which will add its contribution.
As indicated in the introduction, the finding of He-enhancement in  the CRUMPS will also
modify the estimates of the mass of these objects.
It may be that the most massive of these He-rich stars now populate the blue
end of the horizontal branch.
 
If He-abundance is found to be normal, namely near the primordial helium abundance, then
it favors either models invoking small amount of accreted mass (as in the AGB model described above),
or models with large dilution factors.  
In that last case the very low abundance of Li observed in some
CEMP stars should be due to depletion mechanism in the star itself. 
Many of the spinstar models discussed here are compatible with normal He-stars (see Table \ref{models}).
In particular, 
the wind of spinstars model would remain an interesting model for normal He-stars showing
small  $^{12}$C/$^{13}$C number ratios, 
since combinations of dilution factor and of the initial mass of the spinstar may
yield very modest helium enrichments, while still allowing very low $^{12}$C/$^{13}$C number ratios. 

\acknowledgements{The authors are very grateful to Dr Takuma Suda for giving access to the very useful
SAGA database  (http://www.astro.keele.ac.uk/saga/), to Dr Thomas Masseron for providing before publication the electronic tables of the paper
by \citet{Masseron2009} and to Dr Andrea Dupree for interesting discussions on He lines in cool stars.
R. Hirschi acknowledges support from the Marie Curie grant IIF 221145 and an STFC rolling grant.}

\bibliographystyle{aa}
\bibliography{MyBiblio}

\end{document}